\newif\iffigs\figstrue

\documentclass[12pt]{article}
\iffigs
  \input epsf
\else
\fi

\batchmode
\newfont{\footscrfont}{rsfs10}
  \newfont{\footbbbfont}{msbm10}
  \newfont{\manfont}{manfnt}
\errorstopmode

\newif\ifscrf\scrftrue
\ifx\footscrfont\nullfont
  \scrffalse
\fi

\newif\ifamsf\amsftrue
\ifx\footbbbfont\nullfont
  \amsffalse
\fi

\def\ppnumber{\vbox{\baselineskip14pt\hbox{CU-TP 967}
\hbox{hep-th/0002004}}}
\def\ppdate{February 2000}
\def\pplogo{\vbox{\kern-\headheight\kern -15pt
\halign{##&##\hfil\cr&{
\ppnumber}\cr\rule{0pt}{2.5ex}&\ppdate\cr}
}}

\makeatletter
\date{}
\def\dedicatory#1{\def\@date{\normalsize\it#1}}
\def\subjclass#1{\def\@thefnmark{}\@footnotetext{1991
    {\it Mathematics Subject Classification.} #1}}
\def\keywords#1{\def\@thefnmark{}\@footnotetext{
    {\it Key words and phrases.} #1}}

\def\ps@firstpage{\ps@empty \def\@oddhead{\hss\pplogo}%
  \let\@evenhead\@oddhead 
}
\def\maketitle{\par
 \begingroup
 \def\thefootnote{\fnsymbol{footnote}}
 \def\@makefnmark{\hbox
 to 0pt{$^{\@thefnmark}$\hss}}
 \if@twocolumn
 \twocolumn[\@maketitle]
 \else \newpage
 \global\@topnum\z@ \@maketitle \fi\thispagestyle{firstpage}\@thanks
 \endgroup
 \setcounter{footnote}{0}
 \let\maketitle\relax
 \let\@maketitle\relax
 \gdef\@thanks{}\gdef\@author{}\gdef\@title{}\let\thanks\relax}

\def\abstract{\if@twocolumn
\section*{Abstract}
\else \small
\begin{center}
{\bf ABSTRACT}
\end{center}
\quotation
\fi}

\def\thebibliography#1{\section*{References\@mkboth
 {REFERENCES}{REFERENCES}}\small\list
 {[\arabic{enumi}]}{\settowidth\labelwidth{[#1]}\leftmargin\labelwidth
 \advance\leftmargin\labelsep
 \usecounter{enumi}}
 \def\newblock{\hskip .11em plus .33em minus .07em}
 \sloppy\clubpenalty4000\widowpenalty4000
 \sfcode`\.=1000\relax}

\newif\iffn\fnfalse

\@ifundefined{reset@font}{\let\reset@font\empty}{} 
\long\def\@footnotetext#1{\insert\footins{\reset@font\footnotesize
    \interlinepenalty\interfootnotelinepenalty
    \splittopskip\footnotesep
    \splitmaxdepth \dp\strutbox \floatingpenalty \@MM
    \hsize\columnwidth \@parboxrestore
   \edef\@currentlabel{\csname p@footnote\endcsname\@thefnmark}\@makefntext
    {\rule{\z@}{\footnotesep}\ignorespaces
      \fntrue#1\fnfalse\strut}}}

\makeatother

\ifamsf
  \newfont{\bigbbbfont}{msbm10 scaled\magstep2}
  \newfont{\bbbfont}{msbm10 scaled\magstep1}  
  \newfont{\smallbbbfont}{msbm8}
  \newfont{\tinybbbfont}{msbm6}
  \newfont{\smallfootbbbfont}{msbm7}
  \newfont{\tinyfootbbbfont}{msbm5}
  \newfont{\biggthfont}{eufm10 scaled\magstep2}
  \newfont{\gthfont}{eufm10 scaled\magstep1}  
  \newfont{\smallgthfont}{eufm8}
  \newfont{\tinygthfont}{eufm6}
  \newfont{\footgthfont}{eufm10}
  \newfont{\smallfootgthfont}{eufm7}
  \newfont{\tinyfootgthfont}{eufm5}
\fi

\ifscrf
  \newfont{\scrfont}{rsfs10 scaled\magstep1}  
  \newfont{\smallscrfont}{rsfs7}
  \newfont{\tinyscrfont}{rsfs7}
  \newfont{\smallfootscrfont}{rsfs7}
  \newfont{\tinyfootscrfont}{rsfs7}
\fi

\ifamsf
  \newcommand{\Bbb}[1]{\iffn
      \mathchoice{\mbox{\footbbbfont #1}}{\mbox{\footbbbfont #1}}
      {\mbox{\smallfootbbbfont #1}}{\mbox{\tinyfootbbbfont #1}}\else
      \mathchoice{\mbox{\bbbfont #1}}{\mbox{\bbbfont #1}}
      {\mbox{\smallbbbfont #1}}{\mbox{\tinybbbfont #1}}\fi}
  
\else
  \def\bigbbbfont{\bf}
  \def\Bbb{\bf}
  
\fi

\ifscrf
  \newcommand{\Scr}[1]{\iffn
    \mathchoice{\mbox{\footscrfont #1}}{\mbox{\footscrfont #1}}
    {\mbox{\smallfootscrfont #1}}{\mbox{\tinyfootscrfont #1}}\else
    \mathchoice{\mbox{\scrfont #1}}{\mbox{\scrfont #1}}
    {\mbox{\smallscrfont #1}}{\mbox{\tinyscrfont #1}}\fi}
\else
  \def\Scr{\cal}
\fi

\def\tablerule{\noalign{\hrule}}

\def\C{{\Bbb C}}
\def\F{{\cal F}}

\def\P{{\Bbb P}}

\def\R{{\Bbb R}}
\def\Z{{\Bbb Z}}

\def\W{{\Bbb W}}

\def\bearray{\begin{eqnarray}}
\def\eearray{\end{eqnarray}}
\def\bearraynn{\begin{eqnarray*}}
\def\eearraynn{\end{eqnarray*}}
\def\bfig{\begin{figure}}
\def\efig{\end{figure}}

\def\opeq#1{\advance\lineskip#1 \advance\baselineskip#1
        \advance\lineskiplimit#1}

\def\cM{{\Scr M}}

\def\cD{{\Scr D}}

\def\cMc{{\hfuzz=100cm\hbox to 0pt{$\;\overline{\phantom{X}}$}\cM}}
\def\barcD{{\hfuzz=100cm\hbox to 0pt{$\;\overline{\phantom{X}}$}\cD}}

\def\F#1#2{{}_{#1}F_{#2}}

\ifamsf

\else

\fi

\def\boldone{\relax{\rm 1\kern-.35em 1}}

\newtheorem{Proposition}{Proposition}[section]

\newtheorem{Theorem}{Theorem}[section]
\newtheorem{Lemma}{Lemma}[section]
\newtheorem{Corrolary}{Corrolary}[section]

\newcommand{\be}{\begin{equation}}
\newcommand{\ee}{\end{equation}}
\newcommand{\bea}{\begin{eqnarray}}
\newcommand{\eea}{\end{eqnarray}}

\newcommand{\bp}{\begin{Proposition}}
\newcommand{\ep}{\end{Proposition}}
\newcommand{\bt}{\begin{Theorem}}
\newcommand{\et}{\end{Theorem}}
\newcommand{\bl}{\begin{Lemma}}
\newcommand{\el}{\end{Lemma}}
\newcommand{\bc}{\begin{Corrolary}}
\newcommand{\ec}{\end{Corrolary}}
\newcommand{\nn}{\nonumber}

\def\HG_symbol[#1,#2]{
\left(
\begin{array}{c}
#1\\
#2
\end{array}
\right)
}

\def\foo[#1][#2,#3,#4,#5]{{\rm \bf #1}
\left(
\begin{array}{ccc}
#2&~~~~&#3\\
~~&~~~~&~~ \\
#4&~~~~&#5
\end{array}
\right)
}

\def\Foo[#1][#2,#3]{{\rm \bf #1}
\left(
\begin{array}{c}
#2\\
#3
\end{array}
\right)
}

\def\D[#1,#2][#3,#4]{{\rm D}^{#1}_{#2}
\left(
\begin{array}{c}
#3\\
#4
\end{array}
\right)
}

\def\I[#1,#2,#3,#4](#5){{\rm \bf I}
\left(
\begin{array}{ccc}
#1&~~~~&#2\\
~~&~~~~&~~ \\
#3&~~~~&#4
\end{array}
\right)
(#5)}

\def\M[#1,#2,#3,#4](#5){{\rm \bf G}
\left(
\begin{array}{ccc}
#1&~~~~&#2\\
~~&~~~~&~~ \\
#3&~~~~&#4
\end{array}
\right)
(#5)}

\def\G[#1,#2]{{\rm \Gamma}
\left(
\begin{array}{c}
#1\\
#2
\end{array}
\right)}

\def\B[#1,#2][#3,#4,#5,#6](#7){
\frac{1}{2\pi i}\int_{#1}{d#2~\G[{#3,#4},{#5,#6}]{#7}^{#2}}
}

\def\F[#1,#2][#3,#4](#5){_{#1}{\rm F}_{#2}
\left(
\begin{array}{c}
#3\\
#4
\end{array}
\right)
(#5)}

\usepackage{graphics}

\begin{document}

\title{Collapsing D-branes in one-parameter models and small/large radius 
duality}

\author{C.~I.~Lazaroiu$^{1}$}


\maketitle

\vbox{
\centerline{Department of Physics}
\centerline{Columbia University}
\centerline{New York, N.Y. 10027}
\medskip
\medskip
\bigskip
}

\abstract{We finalize the study of collapsing D-branes 
in one-parameter models 
by completing the analysis of the associated hypergeometric hierarchy. 
This brings further evidence that the phenomenon of collapsing 
6-branes at the mirror of the `conifold' point in IIA compactifications on 
one-parameter Calabi-Yau manifolds is generic. It also completes the reduction 
of the study of higher periods in one-parameter models to a few families which 
display characteristic behaviour. One of the models we consider displays 
an exotic form of small-large radius duality, which is a consequence 
of an ``accidental''  discrete symmetry of 
its moduli space. We discuss the implementation of 
this symmetry at the level of the associated type II string 
compactification and its action on D-brane states. 
We also argue that this model admits two special Lagrangian fibrations 
and that the symmetry can be understood as their exchange. 
}

\vskip .6in

$^1$ lazaroiu@phys.columbia.edu

\pagebreak

\section*{Introduction}

Recent work on ``D-brane geometry'' \cite{D_geometry} 
has 
lead to renewed interest in the quantum analogue of the notion of `size'. A 
necessary preliminary of analyses such as \cite{Douglas_quintic} is the 
identification of those D-brane states which become massless at special 
points in the moduli space of a type II compactification on a Calabi-Yau 
manifold, which in the language of \cite{quantum_volumes,branes} amounts to 
identifying the cycles which acquire zero quantum volume at such a point. 
Many basic questions in quantum geometry still await an answer, 
one of the most important among these being the central issue of marginal 
stability and its implications for the extension of mirror symmetry to 
the the D-brane sector of compactified string theory. Such an extension 
holds promise of 
providing a tool for understanding quantum corrections to the moduli space 
of type IIA BPS saturated D-branes. Progress along these lines 
should enable us to understand the tantalizing conjectures of 
\cite{kontsevich} and \cite{vafa_mirror}. 

One of the obstacles to a  detailed and reasonably general 
investigation of D-brane effects in $N=2$ string compactifications 
is the difficulty 
of performing computations of a basis of periods of the holomorphic 3-form 
{\em throughout} the complex structure 
moduli space of a given Calabi-Yau manifold. In \cite{branes}, we took 
a few steps towards removing this obstacle, at least in the 
one-parameter case, by showing how the largely overlooked
\footnote{An example in which Meijer functions were used for performing 
the analytic continuation of periods can be found in \cite{Zaslow}. 
We thank Erik Zaslow for bringing this reference to our attention.}
but classical technique of Meijer functions \cite{Meijer_refs, Norlund} 
can be used to 
give a systematic approach to the 
problem. In fact, this technique allows us to 
reduce most one-parameter models to four classes, each of which allows  
for universal expressions of a special set of periods introduced in 
\cite{branes}. Determining the analytic continuation of periods for 
all such classes amounts to a complete solution of the problem --- given 
a one parameter model, all that remains to be done is to substitute in these 
expressions for 
the specific values of the hypergeometric parameters. 
In \cite{branes} we made use of this approach in order to undertake 
a systematic study of quantum volumes in one parameter models and along 
a special sub-locus of a two-parameter example. Considerations of space 
prevented us from giving a complete discussion of all
classes of one-parameter models. 
The present paper remedies this lack of completeness by carrying through 
a similar 
analysis  for the last two classes of this hierarchy, which in a certain sense 
are the most degenerate situations. This allows us to bring further evidence 
that the phenomenon noticed in \cite{quantum_volumes,PS} 
of collapsing 6-branes at the mirror of the conifold point is generic
in one-parameter models, and not limited to the case of the quintic
\cite{quintic}, where it 
was first observed.

The last part of the 
paper is concerned with a special example which 
exhibits some rather exotic features. This is a one-parameter family of 
Calabi-Yau complete intersections in seven-dimensional projective space, 
which belongs to the most ``degenerate'' family in our classification. 
As noticed a while ago \cite{Candelas_periods}, the moduli space of 
this model admits a $\Z_2$ symmetry which identifies the small and large 
radius limits. This lead to suspicions 
\cite{geometric_interpretation} that the model  
provides a Calabi-Yau example 
of small-large radius duality. This would give an example of a `T-dual' 
string compactification with reduced (N=2) supersymmetry, with potentially 
interesting implications for phenomenology. Our knowledge 
of a basis of periods allows us to address 
some of the puzzles concerning this model. While doing so in 
Section 3, we will meet with some surprises. Indeed, we will be able to confirm
the suspicions of \cite{geometric_interpretation}, but in a rather unexpected 
way: while small-large radius duality 
is indeed an exact feature of the model, its 
realization involves a certain rotation in the space of states, as well 
as a (less surprising) rescaling of the correlation functions. This 
conclusion, which can be extracted from the direct computation of periods 
by a a careful consideration of branch cuts, 
has some interesting implications for the action of the symmetry 
on the D-brane 
states. In particular, the duality exchanges D2 and D4-branes in the 
mirror, type IIA compactification. In Section 5, we propose an explanation of 
this phenomenon by making use of the ideas of Strominger, Yau and Zaslow 
\cite{SYZ}. We will argue that the model admits {\em two} $T^3$ fibrations, 
which are interchanged by our symmetry. The nontrivial action 
on D2/D4 branes appears as a consequence of the fact that the 
dimension of the holomorphic 
cycle wrapped by the mirror of a given type IIB D-brane 
depends on the position of the original special Lagrangian cycle with 
respect to the fibration: when changing the fibration, the 
interpretation of the mirror state is modified. The existence of this symmetry 
has other interesting implications for the D-brane physics of this model. 
In particular, there exists a two-dimensional space of D-brane states which 
vanish at the mirror of the ``conifold'' point (modulo issues of marginal 
stability). Such states can be interpreted as composites of D4 and D6 branes.

\section{Quantum notions of ``size''}

The problem of understanding the correct string-theoretic generalization 
of the notion of size was considered in \cite{small_distances1,
small_distances2,quantum_volumes} (see also \cite{branes} for a review). 
The best framework for addressing this issue is that of type II string 
compactifications on Calabi-Yau manifolds, which have the advantage of 
allowing for exact computations of stringy corrections while at the same time 
displaying nontrivial quantum effects. 
This problem can be approached by considering a type 
IIA compactification on a Calabi-Yau manifold $X$ and its dual, type IIB 
compactification on the mirror $Y$ of $X$. The 
quantum corrections to the notion of size appear in the vector multiplet 
moduli space, which corresponds to the Kahler moduli of the IIA 
compactification and to the complex structure moduli of its IIB dual. 
Since the latter does not suffer quantum  
corrections \cite{strominger_vm}, 
one can use mirror symmetry in order to transport the results accessible 
on this side to the IIA compactification, thereby extracting exact information 
about the stringy corrections to the Kahler moduli space of $X$. Hence mirror 
symmetry identifies the quantum-corrected complexified Kahler moduli space 
of $X$ with the complex structure moduli space ${\cal M}$ of $Y$.

The first question one encounters in this framework is that of introducing 
a physically meaningful parameterization of the corrected
complexified Kahler moduli 
space, which will allow us to measure `quantum areas' 
on $X$. In this paper, we will follow the proposal of \cite{small_distances1}, 
which consists of using the value of the complexified Kahler class dictated 
by the mirror map:
\be
\label{mirror_map}
k(z)=(B+iJ)(z)=\frac{\int_{\gamma_1}\Omega(z)}{\int_{\gamma_0}\Omega(z)}~~,
\ee\noindent 
where $z$ is a coordinate\footnote{We restrict to one-parameter models for 
simplicity.} on ${\cal M}$, 
$\Omega$ is the holomorphic 3-form of $Y$ and $\gamma_0$, $\gamma_1$ are 
certain 3-cycles in $Y$ which can be identified in the manner discussed 
in \cite{quintic, morrison_quintic,morrison_fuchs, morrison_cpctfs,
morrison_aspects}. 
Hence (\ref{mirror_map}) defines a specific 
class in $H^2(X,\C)$ at each point $z$, which is
identified as the correct quantum counterpart of the complexified Kahler class 
at that point. The imaginary part $J$ of this class defines the so-called 
`nonlinear sigma model measure' on ${\cal M}$. More precisely, writing:
\be
k(z)=t(z)e~~,
\ee\noindent where $e$ is the generator of $H^2(X,\Z)$ defines a  
special coordinate on ${\cal M}$ (in the sense of special geometry). Then 
the nonlinear sigma model measure is defined by the imaginary part of $t(z)$.

An ``intermediate'' parameterization of ${\cal M}$ 
is given by the so-called ``algebraic 
coordinate'', which is defined through:
\be
\label{alg_measure}
k_{alg}=(b+is)(z)=\frac{1}{2\pi i}\log (\kappa z)e~~,
\ee\noindent where $\kappa$ is a certain constant which is determined 
by the monomial -divisor mirror map of \cite{mdmm}. 
Measuring distances with $k_{alg}$ amounts to using the semiclassical notion 
of size (which is, strictly speaking, only valid in the large radius limit 
of $X$) throughout the entire moduli space ${\cal M}$. 

An important point, first noticed in \cite{quantum_volumes} and discussed 
in full generality in \cite{branes} is that the classical geometric relation:
\be 
\label{classical_volume}
{\rm vol}(\Sigma_{2p})\sim \int_{\Sigma_{2p}}{k^p}
\ee 
(with $\Sigma_{2p}$ some 2p-cycle in $X$) does not admit a natural 
generalization to the quantum level. This follows by noticing that the 
most natural extension of the notion of volume to the quantum setting is 
to identify the ``quantum volume'' of $\Sigma_{2p}$ with the mass of a $D_{2p}$
brane wrapping this cycle (divided by the associated D-brane tension). 
This can be computed via mirror symmetry 
techniques in the BPS case (when $\Sigma_{2p}$ is a holomorphic cycle and 
hence the associated D-brane state is BPS), since the mass 
of the mirror state (a type IIB $D3$-brane wrapping a special Lagrangian 
3-cycle $C$ mirror to $\Sigma_{2p}$) is given by the exact formula:
\be
m(C)=\frac{|\int_{C}{\Omega}|}{|\int_{C}{\overline \Omega}\wedge
\Omega|^{1/2}}=m(\Sigma_{2g})~~.
\ee\noindent The disagreement between quantum volumes measured in this way and 
those given by (\ref{classical_volume}) is due to open string instanton 
corrections\footnote{These are induced by open strings whose 
endpoints are constrained to lie in $\Sigma_{2p}$.}
to the mass of the corresponding $D_{2p}$ brane \cite{Ooguri}. 
In fact,  
using the semiclassical relation (\ref{classical_volume}) amounts 
to substituting 
the correct quantum Kahler class into the {\em classical} relation
for volumes---a procedure somewhat akin 
to using the algebraic measure (\ref{alg_measure}) instead of the correct, 
nonlinear sigma model measure.

An important question raised by these considerations is to what extent this 
notion of quantum volume behaves like its geometric counterpart. Since the 
definition discussed above includes nontrivial quantum corrections from open 
string instantons, it is natural to expect that the two quantities will diverge
as we move away from the large radius limit of $X$ into regions of the moduli 
space where such corrections are important. In fact, instanton corrections 
are especially strong in the vicinity of conifold points, so one expects 
that the most pronounced difference will be manifest there. This 
suspicion is  confirmed by the observation of \cite{quantum_volumes,
PS} that the quantum volume of IIA $D2$ and $D4$ branes on the quintic 
remains nonzero at the mirror of the conifold point, while the quantum volume 
of a $D6$-brane vanishes. In \cite{branes}, we presented evidence that this 
is a widespread phenomenon in Calabi-Yau compactifications, and not a 
peculiarity of the quintic. However, the analysis of \cite{branes} was limited 
to only two of the four hypergeometric families of one parameter models. 
The purpose of next three sections is to complete this argument,
by showing that the same behaviour occurs in the remaining families, 
thus providing more evidence that this is 
a generic feature of one-parameter compactifications. 

Most of the results of the next three 
sections are of a somewhat technical nature 
and represent a direct extension of the work of \cite{branes}. The reader 
mainly interested in the discussion of Calabi-Yau  small-large radius 
duality can proceed directly to Section 5.

\section{Universal results for one-parameter models}

This section reviews and completes some results obtained in \cite{branes}, 
which will be used intensively below. These rest on the 
theory of Meijer functions \cite{Meijer_refs,Norlund}, a brief account of 
which can be found in \cite{branes}.

\subsection{Review of large radius results}

Let us start by summarizing some material presented in  \cite{branes}. 
Following the discussion of that paper, we  
focus on one-parameter models whose hypergeometric symbol has the form 
{\footnotesize $\HG_symbol[{\alpha_1,\alpha_2,\alpha_3,\alpha_4},{1,1,1}]$}
with $\alpha_j$ some rational numbers contained in the interval $[0,~1]$. 
In this case, the associated Picard-Fuchs equation has the hypergeometric form:
\be
\label{HG_eq1}
\left[
\delta^4-
z (\delta+\alpha_1)(\delta+\alpha_2)(\delta+\alpha_3)(\delta+\alpha_4)
\right]u=0~~
\ee\noindent (where $\delta:=z\frac{d}{dz}$). By using the theory of Meijer 
functions, it was shown in \cite{branes} that an especially convenient 
basis of periods (called {\em Meijer periods}) 
is given by the integral representations:
\be
U_j(z)=\frac{1}{2\pi i}{\int_{\gamma}{ds~\phi_j(s)}}~~,
\ee\noindent where:
{\footnotesize \bea
\label{phi}
\phi_j(s):=\frac{1}{\prod_{i=1~...~4}{\Gamma(\alpha_i)}}~
\frac{{\Gamma(-s)}^{j+1}\prod_{i=1~...~4}{\Gamma(s+\alpha_i)}}{{\Gamma(s+1)}^{3-j}}((-1)^{j+1}z)^s~~. 
\eea}\noindent In these expressions, the contour $\gamma$ is chosen as shown 
in Figure 1.

\vskip 0.5in
\hskip 1.2in\scalebox{0.4}{\begin{picture}(0,0)%
\epsfbox{contour1_0.pstex}%
\end{picture}%
\setlength{\unitlength}{3947sp}%
\begingroup\makeatletter\ifx\SetFigFont\undefined%
\gdef\SetFigFont#1#2#3#4#5{%
  \reset@font\fontsize{#1}{#2pt}%
  \fontfamily{#3}\fontseries{#4}\fontshape{#5}%
  \selectfont}%
\fi\endgroup%
\begin{picture}(7824,6431)(589,-5805)
\put(7951,-61){\makebox(0,0)[lb]{\smash{\SetFigFont{17}{20.4}{\rmdefault}{\mddefault}{\itdefault} s}}}
\end{picture}
}

\

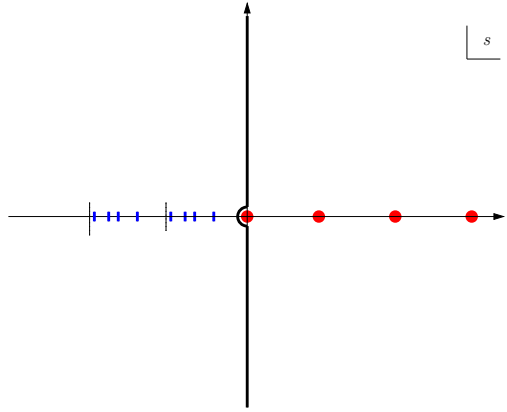
\begin{center}Figure 1. {\footnotesize The 
defining contour for the Meijer periods.}
\end{center}

\

The expansions of these periods in the large and small radius regions of 
the moduli space follow by closing the contour to the right or left, which is 
allowed for $|z|<1$ and $|z|>1$ respectively. The expansions for $|z|<1$ were
computed in \cite{branes} and are given by the universal expression:
\be
\label{LCS_expansions}
U_j(z)=\frac{(-1)^j}{j!}\sum_{n=0}^{\infty}{
\frac{(\alpha_1)_n(\alpha_2)_n(\alpha_3)_n
(\alpha_4)_n}{n!^4}\nu_j(n,z)z^n}~~,
\ee\noindent where:
{\footnotesize 
\bea
\label{nus}
\nu_0&=&1\nn\\
\nu_1(n,z)=g_1'(n,z)&=&\eta_1(n)+{\rm log}(z)\nn\\
\nu_2(n,z)=g_2''(n,z)+[g_2'(n,z)]^2]&=&\eta_2'(n)+(\eta_2(n)+{\rm log}(-z))^2
~~~~~\\
\nu_3(n,z)=g_3'''(n,z)+3g_3''(n,z)g_3'(n,z)+g_3'(n,z)^3\nn &=&
\eta_3''(n)+3\eta_3'(n)(\eta_3(n)+{\rm log}z)+(\eta_3(n)+{\rm log}z)^3~~,
\eea}\noindent with:
{\footnotesize
\be
\label{etas}
\eta_j^{(i)}(n)=
\sum_{k=1}^{4}{\psi^{(i)}(n+\alpha_k)}-(3-j)\psi^{(i)}(n+1)
-(-1)^i(j+1)\left[\psi^{(i)}(1)+i!\sum_{l=1}^{n}\frac{1}{l^{i+1}}
\right]~~,
\ee}\noindent for $i=0,1,2$. In \cite{branes}, we also computed the monodromy 
matrix of the Meijer periods about the large complex structure point $z=0$, 
with the result:
{\footnotesize
\be
T[0]=\left[\begin {array}{cccc} 
1&0&0&0\\
-2i\pi &1&0&0\\
-4\pi^2&-2i\pi &1&0\\
0&0&-2i\pi &1
\end {array}\right ]~~.
\ee}

\subsection{The special coordinate on the moduli space}

For later use, let us derive a universal expression for the special 
coordinate $t$ on the moduli space. As explained in \cite{morrison_quintic, 
morrison_cpctfs,morrison_fuchs,morrison_aspects, mdmm}, 
this is given by a certain 
ratio of a linear 
combination of $\log^0$ and $\log^1$ periods to a $\log^0$ period (the latter 
is, of course, determined up to a global factor). The correct 
linear combination appearing 
in the numerator is fixed by the requirement that the 
asymptotic form of $t$ in the large complex structure limit be given by:
\be
t_{as}=\frac{1}{2\pi i}\log w~,
\ee  
where $w=\kappa z$, with $\kappa=e^{\sum_{k=1}^4{\psi(\alpha_k)}-4\psi(1)}$, 
is a coordinate on the moduli space determined by the monomial-divisor 
mirror map of \cite{mdmm}. The asymptotic form of the Meijer
periods at large complex structure 
can be easily extracted from the expansions in terms of $w$ 
computed in \cite{branes}. Indeed, it was shown there that  
(\ref{LCS_expansions}) can be rewritten as:
\be
U_j(w)=\sum_{s=0}^{j}{{\tilde q}_{sj}(w)({\rm log} w)^s}~~,
\ee
where:
\be
{\tilde q}_{sj}(w):=\frac{(-1)^j}{j!}\sum_{n=0}^{\infty}
{\frac{(\alpha_1)_n(\alpha_2)_n(\alpha_3)_n(\alpha_4)_n}{n!^4}
{\tilde v}_{sj}(n)\left(\frac{w}{\kappa}\right)^n}~~, 
\ee
with ${\tilde v}_{sj}(n)$ some quantities whose explicit 
form is listed in Subsection 
4.2.1 of \cite{branes}. Since the matrix 
${\tilde q}(0):=(q_{sj}(0))_{s,j=0..3}$ has a finite limit at $w=0$,
{\footnotesize
\be
{\tilde q}(0):=
\left[\begin{array}{cccc} 
1&0&\frac{1}{2}(\eta_2'(0)-\pi^2)&-\frac{1}{6}\eta_3''(0)\\
0&-1& i\pi &-\frac{1}{2}\eta_3'(0)\\
0&0&\frac{1}{2}&0\\
0&0&0&-\frac{1}{6}\\
\end{array}\right]~~,
\ee}\noindent it follows that the leading terms in the large radius expansions 
of the Meijer periods are:
\be
U_j^{as}(w)=\sum_{s=0}^{j}{{\tilde q}_{sj}(0)({\rm log} w)^s}~~. 
\ee
In particular, we obtain:
\be
U_0^{as}(w)=1~~,~~U_1^{as}(w)=-\log w
\ee \noindent 
and since the periods $U_j$ are adapted to the monodromy weight filtration of 
the model we immediately deduce that the special coordinate has the simple  
universal form:
\be
\label{special}
t=-\frac{1}{2\pi i}\frac{U_1}{U_0}~~.
\ee \noindent Substituting expansion (\ref{LCS_expansions}) in this 
formula leads to a general expression for the special coordinate in the 
large radius region $|z|<1$ (which can be used, in particular, to extract 
{\em universal} expressions for the 
Gromov-Witten invariants \cite{gromov, witten_tsm} 
of this class of models as functions of the 
parameters $\alpha_k$). On the other hand, 
the analytic continuations of $U_0$ and $U_1$ allow us to compute $t$ 
as a function of $z$ (or $w$) throughout the moduli space.

\subsection{The hypergeometric hierarchy}

As discussed in \cite{branes}, the nature of the small radius expansions of 
the Meijer periods, 
and hence the nature of the small radius point of the model, 
depend on the relative values of the parameters $\alpha_i$. From an abstract 
point of view, this leads to a hierarchy of models characterized (up to 
permutations of $\alpha_i$) by one of the conditions:

\

$(0)$ all $\alpha_i$ are distinct

$(1)$ three of the parameters $\alpha_i$ are distinct

$(2)$ $\alpha_1=\alpha_2$ and $\alpha_3=\alpha_4$ but $\alpha_1\neq \alpha_3$

$(3)$ $\alpha_1=\alpha_3=\alpha_3=\alpha_4$.

$(4)$ $\alpha_1=\alpha_2=\alpha_3\neq \alpha_4$

\

\noindent Only levels $(0),~(1),~(2)$ and $(3)$ of this hierarchy are realized 
through one-parameter complete intersections in projective spaces, as well as 
through many one-parameter complete intersections in weighted projective 
spaces and more general toric varieties (see \cite{toric} for a discussion of 
toric geometry).
Level $(4)$ does not seem to be realized\footnote{This follows from 
the results of \cite{batyrev_straten}.} through compact 
one-parameter complete 
intersections in toric varieties, though it could be realized  through more 
general constructions. Since we are mostly interested in the toric case, we 
will limit ourselves to the families (0),~(1),~(2) and (3). 
A few examples of models belonging to these classes are listed in 
Table 1.

\

{\scriptsize \medskip \centerline{
$$\vbox{\offinterlineskip \tabskip=0pt
\halign{#&
\vrule height 14pt depth 7pt
\enskip\hfil$#$\hfil\enskip\vrule &
\enskip\hfil$#$\hfil\enskip\vrule &
\enskip\hfil$#$\hfil\enskip\vrule \cr\tablerule &
Family & Model &~(\alpha_1,~\alpha_2,~\alpha_3,~\alpha_4) \cr\tablerule &
0 &\P^4[5] &(1/5,~2/5,~3/5,~4/5) \cr\tablerule &
0 &\W\P^{2,1,1,1,1}[6] &(1/3,~2/3,~1/6,~5/6) \cr\tablerule &
0 &\W\P^{4,1,1,1,1}[8] &(1/8,~3/8,~5/8,~7/8) \cr\tablerule &
0 &\W\P^{5,2,1,1,1}[10] &(1/10,~3/10,~7/10,~9/10) \cr\tablerule &
0 &\W\P^{2,1,1,1,1,1}[3,4] &(1/3,~2/3,~1/4,~3/4) \cr\tablerule &
0 &\W\P^{3,2,2,1,1,1}[4,6] &(1/6,~1/4,~3/4,~5/6) \cr\tablerule &
1 &\P^5[2,4] &(1/2,~1/2,~1/4,~3/4) \cr\tablerule &
1 &\P^6[2,2,3] &(1/2,~1/2,~1/3,~2/3) \cr\tablerule &
1 &\W\P^{3,1,1,1,1,1}[2,6] &(1/2,~1/2,~1/6,~5/6) \cr\tablerule &
2 &\P^5[3,3] &(1/3,~1/3,~2/3,~2/3) \cr\tablerule &
2 &\W\P^{2,2,1,1,1,1}[4,4] &(1/4,~1/4,~3/4,~3/4) \cr\tablerule &
2 &\W\P^{3,3,2,2,1,1} &(1/6,~1/6,~5/6,~5/6) \cr\tablerule &
3 &\P^7[2,2,2,2] &(1/2,~1/2,~1/2,~1/2) \cr\tablerule 
}}$$ }}

\vskip .3in

\begin{center}
\footnotesize{Table 1. Some examples of models belonging to various 
hypergeometric families.}
\end{center}

\

In \cite{branes}, we studied only the families $(0)$ and $(1)$. 
Here we consider the more degenerate 
families $(2)$ and $(3)$. As we show below, these models can also be 
approached efficiently by the general methods developed in \cite{branes}. 
The highly degenerate family $(3)$ displays some surprising, 
which we discuss in detail in Section 5.

\subsection{The choice of branch-cuts}

Let us clarify the choice of branch-cuts used in the present paper 
and implicitly in \cite{branes}. Our convention is that we start from the 
large complex structure region $|z|<1$ and perform the analytic continuation 
through the 
sector ${\rm arg}(z)\in (-\pi,0)$, i.e. through the lower half of 
the unit circle 
in the complex plane (see Figure 2). Moreover, we will pick the branch-cut 
of all periods to lie along the negative real 
axis $(-\infty,0)$. With this convention, expressions such as $\log(-z)$, 
$\log(-1/z)$ and $\log(z)$ are always understood to have the cut on the 
negative real axis, so that we can write:
\be
\log(-z)=\log(z)+i\pi~~,~~\log(-1/z)=-\log(z)-i\pi~~.
\ee\noindent A similar convention is used for power functions with non-integral 
exponents. In particular, we have 
$(-z)^{-\alpha}=z^{\alpha}e^{-i\pi \alpha}$ for any real constant $\alpha$. 
For the `generic' model considered in \cite{branes} 
the branch-cut along $(-\infty, 0)$ suffices for all periods. 
For the other families (and in particular for all
models discussed  in the present paper), the situation is slightly different 
since in these cases the analytic 
continuation of the fundamental period $U_0$ displays logarithmic behaviour 
in the region $|z|>1$, even though it is regular in the unit disk, $|z|<1$. 
This requires that we enlarge the associated branch-cut in a way 
consistent with this behaviour, and we shall do so by adding the upper half of 
the unit circle to the common cut along the negative real axis.

\vskip 0.5in 
\hskip 1.2in\scalebox{0.4}{\input{cuts.pstex_t}}
\vskip 0.3in
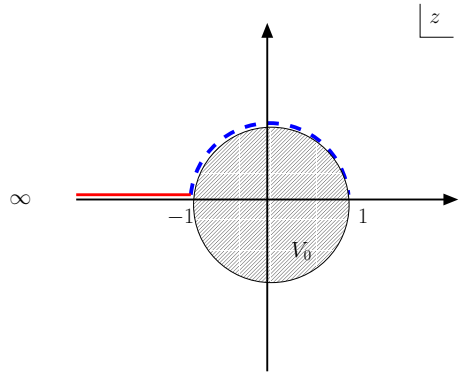
\hskip 1in \begin{center}Figure 2. {\footnotesize Our choice of branch-cuts 
for the analytic continuation of periods. The upper half of the unit circle 
is added only for the fundamental period $U_0$, in all cases when this period
displays logarithmic behaviour in the region $|z|>1$.}
\end{center}

\section{The family $\alpha_1=\alpha_2$,~$\alpha_3=\alpha_4$}

Consider first the family $(2)$, which corresponds to 
the hypergeometric symbol {\footnotesize 
$\HG_symbol[{\alpha,\alpha,\beta,\beta},{1,1,1}]$}, i.e. to the parameters
$\alpha_1=\alpha_2:=\alpha$, $\alpha_3=\alpha_4:=\beta$ with $\alpha\neq \beta$,
where we take $0<\alpha,~\beta<1$.  

\subsection{The Meijer periods}

The expansion of the Meijer periods for $|z|<1$ follows by substituting our 
particular values for $\alpha_i$ in the general formula 
(\ref{LCS_expansions}). The expansions for $|z|>1$ follow by closing the 
contour to the right, which gives contributions from the B-type poles:

\

$(B_1)$~$s=-n-\alpha$

\

$(B_2)$~$s=-n-\beta$

\

\noindent (with $n$ a nonnegative integer). Noticing that all such poles 
are double, a straightforward residue computation yields:
{\footnotesize
\bea
\label{LG_expansions3}
U_j(z)=\left(\frac{\sin \pi \alpha}{\pi} \right)^{3-j}((-1)^{j+1}z)^{-\alpha}
\sum_{n=0}^{\infty}
{\frac{\Gamma(n+\alpha)^4\Gamma(-n+\beta-\alpha)^2}{\Gamma(\alpha)^2\Gamma(\beta)^2 n!^2}z^{-n}}~\times~~~~~~~~~~~~~~~~~~~~~~~~~~~~~~~~~~~~~~~~~~~~\nn\\
\left[2\psi(1)+2\psi(-n+\beta-\alpha)-(j+1)\psi(n+\alpha)-(3-j)
\psi(-n-\alpha+1)+2\sum_{k=1}^{n}{\frac{1}{k}}+\log((-1)^{j+1}z)
\right]~~~~~~~~~~~~~~~~\\
+(\alpha\longleftrightarrow\beta)~~.~~~~~~~~~~~~~\nn
\eea}

\subsection{Meijer monodromies}

The monodromy of the Meijer basis about $z=0$ follows by applying the  
results reviewed above, while the monodromy about $z=\infty$
can be computed by making use of the 
general techniques developed in \cite{branes}. Following that procedure, 
we first determine the 
canonical and Jordan forms of the matrix $R[\infty]$:
{\footnotesize 
\bea
R_{can}[\infty]=
\left [\begin {array}{cccc} 0&-1&0&0\\0&0&-1&0
\\0&0&0&-1\\{\alpha}^{2}{\beta}^{2
}&-2{\alpha}^{2}\beta-2\alpha{\beta}^{2}&{\alpha}^{2}+4\alpha
\beta+{\beta}^{2}&-2\alpha-2\beta\end {array}\right ]~~,~~
R_J[\infty]=\left [\begin {array}{cccc} 
-\beta&1&0&0\\0&-\beta&0&0\\0&0&-\alpha&1\\0&0&0&-\alpha
\end {array}\right ]~~.\nn
\eea}\noindent The relation $R_{can}[\infty]=PR_J[\infty]P^{-1}$  allows us to 
determine a choice for the transition matrix $P$ from a Jordan basis to the 
canonical basis:
{\footnotesize 
\bea
P=\left [\begin {array}{cccc} {\frac {{\alpha}^{2}\beta}{{\alpha}^{2}-2
\alpha\beta+{\beta}^{2}}}&{\frac {\left (\alpha-3\beta\right ){
\alpha}^{2}}{{\alpha}^{3}-3{\alpha}^{2}\beta+3\alpha{\beta}^{2}-
{\beta}^{3}}}&{\frac {\alpha{\beta}^{2}}{{\alpha}^{2}-2\alpha
\beta+{\beta}^{2}}}&{\frac {\left (3\alpha-\beta\right ){\beta}^{2}}
{{\alpha}^{3}-3{\alpha}^{2}\beta+3\alpha{\beta}^{2}-{\beta}^{3}}
}\\{\frac {{\alpha}^{2}{\beta}^{2}}{{\alpha}^{2}-2
\alpha\beta+{\beta}^{2}}}&-2{\frac {{\alpha}^{2}{\beta}^{2}}{{
\alpha}^{3}-3{\alpha}^{2}\beta+3\alpha{\beta}^{2}-{\beta}^{3}}}&
{\frac {{\alpha}^{2}{\beta}^{2}}{{\alpha}^{2}-2\alpha\beta+{\beta}
^{2}}}&2{\frac {{\alpha}^{2}{\beta}^{2}}{{\alpha}^{3}-3{\alpha}^{2
}\beta+3\alpha{\beta}^{2}-{\beta}^{3}}}\\{\frac 
{{\beta}^{3}{\alpha}^{2}}{{\alpha}^{2}-2\alpha\beta+{\beta}^{2}}}&
-{\frac {{\alpha}^{2}{\beta}^{2}\left (\alpha+\beta\right )}{{\alpha}^
{3}-3{\alpha}^{2}\beta+3\alpha{\beta}^{2}-{\beta}^{3}}}&{\frac {{\alpha}^{4}{\beta}^{2}}{{\alpha}^{2}-2\alpha\beta+{\beta}
^{2}}}&2{\frac {{\alpha}^{3}{\beta}^{3}}{{\alpha}^{3}-3{\alpha}^{2
}\beta+3\alpha{\beta}^{2}-{\beta}^{3}}}\end {array}\right ]~~.
\eea}\noindent In the present case, the 
singular content of the periods around $z=\infty$ can be extracted by writing 
$U^t(z)=Z(z)q(z)$, where {\footnotesize $Z(z)=\left[\begin{array}{cccc}
z^{-\alpha}&z^{-\alpha}\log z&z^{-\beta}&z^{-\beta}\log z
\end{array}\right]$} and $q(z)=(q_{sj}(z))_{s,j=0..3}$, with:
{\footnotesize
\bea
q_{0j}(z)=\left(\delta_{j,odd}+\delta_{j,even}e^{i\pi\alpha}\right)
\left(\frac{\sin \pi \alpha}{\pi} \right)^{3-j}
\sum_{n=0}^{\infty}
{\frac{\Gamma(n+\alpha)^4\Gamma(-n+\beta-\alpha)^2}{\Gamma(\alpha)^2\Gamma(\beta)^2 n!^2}z^{-n}}~\times~~~~~~~~~~~~~~~~~~~~~~~~~~~~~~~~~~~~~~~\nn\\
\left[2\psi(1)+2\psi(-n+\beta-\alpha)-(j+1)\psi(n+\alpha)-(3-j)
\psi(-n-\alpha+1)+2\sum_{k=1}^{n}{\frac{1}{k}}+i\pi\delta_{j,even})
\right]~~\nn\\
q_{1j}(z)=\left(\delta_{j,odd}+\delta_{j,even}e^{i\pi\alpha}\right)
\left(\frac{\sin \pi \alpha}{\pi} \right)^{3-j}
\sum_{n=0}^{\infty}
{\frac{\Gamma(n+\alpha)^4\Gamma(-n+\beta-\alpha)^2}{\Gamma(\alpha)^2\Gamma(\beta)^2 n!^2}z^{-n}}~~,~~~~~~~~~~~~~~~~~~~~~~~~~~~~~~~~~~~~~~~~~\nn
\eea}
\noindent and $q_{2j}(z)=q_{0j}(z)|_{\alpha\leftrightarrow\beta}$, 
$q_{3j}(z)=q_{1j}(z)|_{\alpha\leftrightarrow\beta}$.

On the other hand, the matrix $z^{R_J[\infty]}$ has the simple form:
{\footnotesize
\be
z^{R_J[\infty]}=
\left [\begin {array}{cccc} {z}^{-\beta}&\ln (z){z}^{-\beta}&0&0
\\0&{z}^{-\beta}&0&0\\0&0&{z}^{-
\alpha}&\ln (z){z}^{-\alpha}\\0&0&0&{z}^{-\alpha}
\end {array}\right ]~~.
\ee}\noindent This allows us to find the matrix $q_J(z)$ which satisfies
$U^t_J(\infty)=Z(z)q_J(z)$:
{\footnotesize
\be
q_J(z)=\left [\begin {array}{cccc} 0&0&S_{{1,3}}(z)&S_{{1,4}}(z)
\\0&0&0&S_{{1,3}}(z)\\S_{{1,1}}(z)&S_{{1
,2}}(z)&0&0\\0&S_{{1,1}}(z)&0&0\end {array}\right ]~~.
\ee}\noindent In this expression, $S_{ij}(z)$ are the entries of the matrix 
$S(z)$ which 
defines the nilpotent orbit of the fundamental system $\Phi_J(z)$ associated 
with the Jordan basis $U_J(z)$:
\be
\Phi_J(z)=S(z)z^{R_J}~~.
\ee
\noindent Since $S(\infty)=P$, we obtain:
{\footnotesize
\be
q_J(\infty)=\left [\begin {array}{cccc} 0&0&{\frac {\alpha{\beta}^{2}}{\left (-\beta+\alpha\right )^{2}}}&{\frac {\left (3\alpha-\beta\right ){
\beta}^{2}}{\left (-\beta+\alpha\right )^{3}}}\\0&0&0
&{\frac {\alpha{\beta}^{2}}{\left (-\beta+\alpha\right )^{2}}}
\\{\frac {{\alpha}^{2}\beta}{\left (-\beta+\alpha
\right )^{2}}}&{\frac {\left (\alpha-3\beta\right ){\alpha}^{2}}{
\left (-\beta+\alpha\right )^{3}}}&0&0\\0&{\frac {{
\alpha}^{2}\beta}{\left (-\beta+\alpha\right )^{2}}}&0&0\end {array}
\right ]~~.
\ee}\noindent We can now compute the matrix 
$M=q(\infty)^tq_J(\infty)^{-t}$ and the Meijer monodromy about the small 
radius point:
\be
T[\infty]=MT_J[\infty]M^{-1}~~,
\ee
where: 
{\footnotesize
\be
T_J[\infty]=e^{2\pi i R_J[\infty]^t}=
\left [\begin {array}{cccc} e^{-2i\pi \alpha}&0&0&0
\\2i\pi e^{-2i\pi \alpha}& e^{-2i\pi \alpha}&0&0\\
0&0& e^{-2i\pi \beta}&0\\
0&0&2i\pi  e^{-2i\pi \beta}& e^{-2i\pi \beta}\end {array}\right ]~~.
\ee}

\subsection{The model $\P^5[3,3]$}

The mirror $Y$ of this model can be realized as an orbifold
\footnote{We refer the reader to \cite{Teitelbaum} for details 
about the orbifold action in this case.} of a complete 
intersection $p_1=p_2=0$ of two cubics in $\P^5$:
\bea
p_1&=&x_1^3+x_2^3+x_3^3-3\psi x_4x_5x_6\nn\\
p_2&=&x_4^3+x_5^3+x_6^3-3\psi x_1x_2x_3 \nn~~.
\eea\noindent The fundamental period and special coordinate in this example
are discussed in \cite{Teitelbaum, Candelas_periods}. Reference 
\cite{Teitelbaum} also discusses the counting of holomorphic curves for this 
model.

In this example, the hypergeometric coordinate is given by 
$z=\frac{1}{\psi^6}$. 
The matrices $R_{can}[\infty], R_J[\infty]$ and a choice for the 
matrix $P$ are given in Appendix A.
This data allows us to compute the Meijer monodromies:
{\footnotesize
\be
T[0]=\left [\begin {array}{cccc} 1&0&0&0\\-2i
\pi &1&0&0\\-4{\pi }^{2}&-2i\pi &1&0
\\0&0&-2i\pi &1\end {array}\right ]~~,~~
T[\infty]=\left [\begin {array}{cccc} 1&0&0&0\\-5&-3{\frac {i}{\pi }}&9/4{\pi }^{-2}&{\frac {9}{8}}{\frac {i
}{{\pi }^{3}}}\\10&15/2{\frac {i}{\pi }}&
-{\frac {27}{4}}{\pi }^{-2}&-9/2{\frac {i}{{\pi }^{3}}}
\\-8&-9{\frac {i}{\pi }}&{\frac {27}{4}}
{\pi }^{-2}&{\frac {27}{4}}{\frac {i}{{\pi }^{3}}}
\end {array}\right ]
\ee} 
and $T[1]=T[0]^{-1}T[\infty]$. These monodromy matrices satisfy:
\be
(T[0]-I)^4=0~~,~~(T[1]-I)^2=0~~,~~(T[\infty]^3-I)^2=0~~\nn~~.
\ee
A set of periods associated with with a basis of a full sublattice of the 
integral lattice $H_3(Y,\Z)$ (up to a {\em common} factor) is given by:
{\footnotesize 
\be
U_E(z)=EU(z)~~,\mbox{with~}~~E=\left [\begin {array}{cccc} 1&0&0&0\\-5&-3{\frac {
i}{\pi }}&9/4{\pi }^{-2}&{\frac {9}{8}}{\frac {i
}{{\pi }^{3}}}\\10&15/2{\frac {i}{\pi }}&
-{\frac {27}{4}}{\pi }^{-2}&-9/2{\frac {i}{{\pi }^{3}}}
\\-8&-9{\frac {i}{\pi }}&{\frac {27}{4}}
{\pi }^{-2}&{\frac {27}{4}}{\frac {i}{{\pi }^{3}}}
\end {array}\right ]~~.
\ee}
It is also easy to check that the period $U_v(z)=\frac{3}{\pi^3}
\left[\frac{3}{8}U_3-\pi^2 U_1\right]$ vanishes at $z=1$. 
This period is weakly integral since:
{\footnotesize
\be
U_v(z)=-i[3,3,2,1]U_E(z)~~.
\ee}\noindent 
The relation $U_v(1)=0$ is equivalent with an arithmetic identity
which we write down in the Appendix. 

In this case, the constant $\kappa=e^{2\psi(\alpha)+2\psi(\beta)-4\psi(1)}=
\frac{1}{729}=3^{-6}$ and the imaginary part of the algebraic coordinate 
on the moduli space is $s=-\frac{1}{2\pi}\log(\kappa |z|)=
\frac{3}{\pi}\log(\frac{3}{|z|})$. Figure 3 displays the values of $|U_v|$ 
versus $s$. The point $z=1$ corresponds to 
$s=\frac{3\log 3}{\pi}\approx 1.049$. For comparison, we also display the 
absolute values of the weakly integral period 
$\frac{9}{4\pi^2}U_2$ and of the special coordinate $t$.  
Figure 4 shows the absolute 
value of the special coordinate $t=-\frac{1}{2\pi i}\frac{U_1}{U_0}$ as a 
function of $s$, for $s\in [-6,2]$. The 
asymptotic form of $t$ in the small radius limit 
$z\rightarrow \infty$ can be easily computed from the small radius expansions 
given above:
{\footnotesize
\bea
t\approx
-{\frac {2(i\sqrt {3}\log (z)+(1+i)\pi)}
{\sqrt{3}\left (-1+i\sqrt {3}\right )\log z}}+
O\left( (\log z)^{-2}\right) 
={\frac {9\log 3-6\pi s-3i\sqrt 
{3}\log 3+i\pi(\sqrt {3} s-2) }{6(-3\log 3+\pi s)}}+O(s^{-2})~~.\nn
\eea}\noindent In particular, the value of $t$ in the limit $z=\infty$ is:
\be
t_{lim}=-\frac{1}{2}+i\frac{\sqrt{3}}{6}\Longrightarrow 
|t_{lim}|=\frac{1}{\sqrt{3}}\approx .577~~.
\ee

\vskip 0.5 in
\begin{center}
$\begin{array}{cc}
\begin{array}{c}\scalebox{0.3}{\begin{picture}(0,0)%
\epsfbox{5_33.pstex}%
\end{picture}%
\setlength{\unitlength}{3947sp}%
\begingroup\makeatletter\ifx\SetFigFont\undefined%
\gdef\SetFigFont#1#2#3#4#5{%
  \reset@font\fontsize{#1}{#2pt}%
  \fontfamily{#3}\fontseries{#4}\fontshape{#5}%
  \selectfont}%
\fi\endgroup%
\begin{picture}(10828,7845)(1243,-8173)
\put(5026,-7561){\makebox(0,0)[lb]{\smash{\SetFigFont{20}{24.0}{\rmdefault}{\bfdefault}{\updefault}$s$}}}
\put(1351,-4336){\makebox(0,0)[lb]{\smash{\SetFigFont{20}{24.0}{\rmdefault}{\bfdefault}{\updefault}$m$}}}
\end{picture}
}\end{array}&
\begin{array}{c}\scalebox{0.39}{\begin{picture}(0,0)%
\epsfbox{5_33t.pstex}%
\end{picture}%
\setlength{\unitlength}{3947sp}%
\begingroup\makeatletter\ifx\SetFigFont\undefined%
\gdef\SetFigFont#1#2#3#4#5{%
  \reset@font\fontsize{#1}{#2pt}%
  \fontfamily{#3}\fontseries{#4}\fontshape{#5}%
  \selectfont}%
\fi\endgroup%
\begin{picture}(9553,6795)(826,-7123)
\put(826,-3511){\makebox(0,0)[lb]{\smash{\SetFigFont{17}{20.4}{\rmdefault}{\bfdefault}{\updefault}$|t|$}}}
\end{picture}
}\end{array}\\
\begin{array}{c}
~\\
\mbox{Figure 3. {\footnotesize  
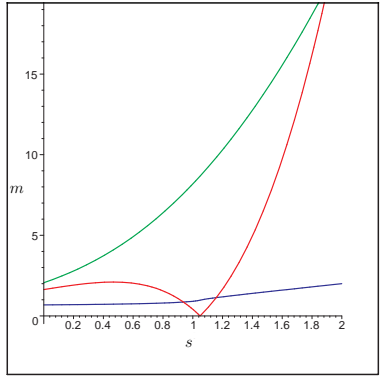
Graph of $|U_v|,~\frac{9}{4\pi^2}|U_2|$ and $|t|$ versus the imaginary }} \\
\mbox{{\footnotesize
part $s$ of the algebraic coordinate for $s \in [0,2]$. The }} \\
\mbox{{\footnotesize
point $z=1$ 
corresponds to $s=\frac{3{\rm log} 3}{\pi}\approx 1.049$. }}\\ 
\end{array}
&
\begin{array}{c}
~~\\
~~\\
\mbox{Figure 4. {\footnotesize  
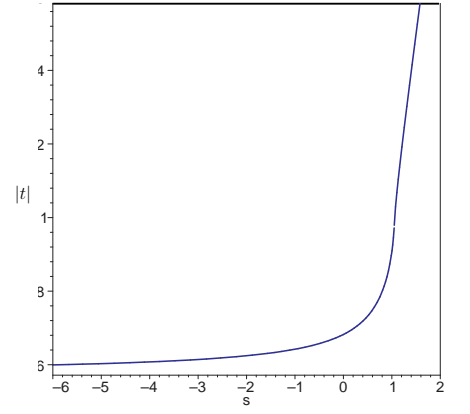
Graph of $|t|$ versus $s$}}\\
\mbox{{\footnotesize for $s \in [-6,2]$.}}\\
~~
\end{array}~~~~~~~~~~~~~
\end{array}$~~~~~~~~~~~~~~~~~~~~~~~~~~~~~~~~~~~~~~~~~~~~~~~~~~~~~~~~~~~~~~~~~~~~~~~~~~~~~~~~~~~~~~~~~~~~~~~~~~~~~~~~~~~~~~~~~~~~~~~~~~~~~~~~~~~~~~~~~~~~~~~~~~~~~~~
\end{center}

\section{The family $\alpha_1=\alpha_2=\alpha_3=\alpha_4$}

In this section we consider the family $(3)$, 
associated with the hypergeometric symbol 
{\footnotesize $\HG_symbol[{\alpha,\alpha,\alpha,\alpha},{1,1,1}]$}, 
i.e. to the parameters $\alpha_1=\alpha_2=\alpha_3=\alpha_4=\alpha$,
where we take $0<\alpha<1$ for simplicity.  The hypergeometric equation of 
this family has the form:
\be
\label{HG_eq2}
\left[\delta^4-z(\delta+\alpha)^4 \right]u=0~~.
\ee

\subsection{The Meijer periods}
The expansion of the Meijer periods for $|z|<1$ follows from the general 
results of Section 2, while the expansion for $|z|>1$ is obtained by 
closing the contour to the right and applying the residue theorem. 
This brings contributions from the quadruple poles $s=-n-\alpha$,
with $n$ a nonnegative integer. In this case, the computation is 
rather similar to that leading to the large radius expansions 
(\ref{LCS_expansions}), giving the result:
{\footnotesize
\bea
\label{LG_expansions4}
U_j(z)=\frac{1}{6}\left(\frac{\sin \pi \alpha}{\pi}\right)^{3-j}
((-1)^{j+1}z)^{-\alpha}\sum_{n=0}^{\infty}{
\left[\frac{(\alpha)_n}{n!}\right]^4\mu_j(n,z)z^{-n}}~~,
\eea}
where the quantities $\mu_j(n,z)$ are defined through:
{\footnotesize
\be
\label{mus}
\mu_j(n,z)=\xi^{''}_j(-n-\alpha)+3\xi^{'}_j(-n-\alpha)
\left(\xi_j(-n-\alpha)+i\pi \delta_{j,even}+\log z\right)+
\left(\xi_j(-n-\alpha)+i\pi \delta_{j,even}+\log z\right)^3~~,
\ee}\noindent with:
{\footnotesize
\be
\label{xis}
\xi_j^{(i)}(-n-\alpha)=4\left[\psi^{(i)}(1) +i!\sum_{k=1}^{n}{\frac{1}
{k^{i+1}}}
\right]-(-1)^i(j+1)\psi^{(i)}(n+\alpha)-(3-j)\psi^{(i)}(1-n-\alpha)~~
\ee}\noindent for $i=0,1,2$. 

\subsection{Meijer monodromies}

The monodromies of the Meijer basis can be extracted by a procedure very 
similar to the one employed above. For the benefit of the reader interested 
in reproducing our computations, let us mention that in this case 
the correct row vector needed for extracting the singular behaviour around 
$z=\infty$ is:
{\footnotesize
\be
Z=\left[\begin{array}{cccc}z^{-\alpha}&z^{-\alpha}\log(z)&
z^{-\alpha}\log(z)^2&z^{-\alpha}\log(z)^3\end{array}\right]~~
\ee}\noindent and that writing $U^t(z)=Z(z)q(z)$ produces a regular 
matrix function $q(z)$ 
whose value $q(\infty)$ at the point of interest has entries:
{\footnotesize
\be
q_{ij}(\infty)=\frac{1}{6}\left(\frac{\sin \pi \alpha}{\pi}\right)^{3-j}
\left(\delta_{j,odd}+\delta_{j,even}e^{-i\pi\alpha}\right)v_{ij}(\infty)~~,
\ee}\noindent where:
{\footnotesize
\bea
v_{0j}(\infty)=\xi_j^{''}(-\alpha)+3\xi^{'}_j(-\alpha)
\left(\xi_j(-\alpha)+i\pi \delta_{j,even}\right)+
\left(\xi_j(-\alpha)+i\pi \delta_{j,odd}\right)^3~~,~~~
v_{3j}(\infty)=1~~.\nn\\
v_{1j}(\infty)=3\xi^{'}_j(-\alpha)+3\left(\xi_j(-\alpha)+
i\pi \delta_{j,even}\right)^2~~~~~~~~,~~~~~~~~
v_{2j}(\infty)=3\left(\xi_j(-\alpha)+i\pi \delta_{j,odd}\right)~~.\nn
\eea}\noindent (Here $\xi^{(i)}_j(-\alpha)$ are obtained from 
(\ref{xis}) by setting $n=0$.)

The canonical and Jordan forms of the matrix $R[\infty]$ are:
{\footnotesize
\be
R_{can}[\infty]=
\left [\begin {array}{cccc} 0&-1&0&0\\0&0&-1&0\\0&0&0&-1\\{\alpha}^{4}&-4{
\alpha}^{3}&6{\alpha}^{2}&-4\alpha\end {array}\right ]
~~,~~
R_J[\infty]=
\left [\begin {array}{cccc} -\alpha&1&0&0\\0&-\alpha
&1&0\\0&0&-\alpha&1\\0&0&0&-\alpha
\end {array}\right ]
\ee}\noindent while a choice for the matrix $P$ which defines a Jordan basis 
is:
{\footnotesize
\be
P=
\left [\begin {array}{cccc} {\alpha}^{3}&{\alpha}^{2}&\alpha&1
\\{\alpha}^{4}&0&0&0\\{\alpha}^{5}
&-{\alpha}^{4}&0&0\\{\alpha}^{6}&-2{\alpha}^{5}&{
\alpha}^{4}&0\end {array}\right ]~~.
\ee}\noindent The matrix $q_J(z)$ is expressed in terms of the nilpotent
orbit $S(z)$ of $\Phi_J(z)$ via:
{\footnotesize
\bea
q_J(z)=\left [\begin {array}{cccc} S_{{1,1}}(z)&S_{{1,2}}(z)&S_{{1,3}}(z)&
S_{{1,4}}(z)
\\0&S_{{1,1}}(z)&S_{{1,2}}(z)&S_{{1,3}}(z)
\\0&0&1/2S_{{1,1}}(z)&1/2S_{{1,2}}(z)
\\0&0&0&1/6S_{{1,1}}(z)\end {array}\right ]~~
\stackrel{\mbox{{\tiny $(S(\infty)=P)$}}}{\Longrightarrow}~~q_J(0)=
\left [\begin {array}{cccc} {\alpha}^{3}&{\alpha}^{2}&\alpha&1
\\0&{\alpha}^{3}&{\alpha}^{2}&\alpha
\\0&0&1/2{\alpha}^{3}&1/2{\alpha}^{2}
\\0&0&0&1/6{\alpha}^{3}\end {array}\right ]~~.\nn
\eea}\noindent Finally, the Meijer monodromy about $z=\infty$ can be computed 
as $T[\infty]=MT_J[\infty]M^{-1}$, where $M=q(0)^tq_J(0)^{-t}$ and:
{\footnotesize
\be
T_J[\infty]=\left[\begin{array}{cccc}
e^{-2\pi i \alpha}&0&0&0\\
2\pi i e^{-2\pi i \alpha}&e^{-2\pi i \alpha}&0&0\\
-2\pi^2e^{-2\pi i \alpha}&2i\pi e^{-2\pi i \alpha}&e^{-2\pi i \alpha}&0\\
-\frac{4i\pi^3}{3}e^{-2\pi i \alpha}&-2\pi^2e^{-2\pi i \alpha}&2\pi i 
e^{-2\pi i \alpha}&e^{-2\pi i \alpha}
\end{array}\right]~~.\nn
\ee}
The expression of the special coordinate (\ref{special}) 
as a function of $z$ follows easily from the small and large radius expansions 
of the Meijer periods. For later reference, we write down the form of 
$t(z)$ in the region $|z|>1$:
{\footnotesize
\be
t(z)=\frac{ie^{i\pi \alpha}}{2\sin \pi \alpha}
\frac{\sum_{n=0}^{\infty}{\left[\frac{(\alpha)_n}{n!}\right]^4
\mu_1(n,z)z^{-n}}}
{\sum_{n=0}^{\infty}{\left[\frac{(\alpha)_n}{n!}\right]^4
\mu_0(n,z)z^{-n}}}~~.
\ee}\noindent This allows us to extract the asymptotic form of $t$ for 
$z\rightarrow \infty$:
{\footnotesize
\be
\label{special_as}
t_{as}=\frac{ie^{i\pi \alpha}}{2\sin \pi \alpha}\frac{\mu_1(0,z)}{\mu_0(0,z)}
= \frac{ie^{i\pi \alpha}}{2\sin \pi \alpha}
\left[1+\frac{3\left(\psi(\alpha)-\psi(1-\alpha)-i\pi\right)}
{\log z}\right]+O\left((\log z)^{-2}\right)~~, 
\ee}\noindent where we used the relations:
{\footnotesize
\bea
\xi_0(-\alpha)=4\psi(1)-\psi(\alpha)-3\psi(1-\alpha)~~,~~
\xi_1(-\alpha)=4\psi(1)-2\psi(\alpha)-2\psi(1-\alpha)~~.\nn
\eea}

\subsection{The model $\P^7[2,2,2,2]$}

The mirror of this model is given by an orbifold $Y$ 
of the complete intersection 
$\{p_1=p_2=p_3=p_4=0\}$, where:
{\footnotesize
\bea
\label{Y}
p_1&=&x_1^2+x_2^2-2\psi x_3 x_4~~\nn\\
p_2&=&x_3^2+x_4^2-2\psi x_5 x_6~~\nn\\
p_3&=&x_5^2+x_6^2-2\psi x_7 x_8~~\\
p_4&=&x_7^2+x_8^2-2\psi x_1 x_2~~.\nn
\eea}\noindent The fundamental period of this example was determined in
\cite{Candelas_periods}( see also \cite{Teitelbaum}), while the 
semiclassical structure of the 
K\"{a}hler moduli 
space was analyzed in detail in \cite{geometric_interpretation} by making use 
of the linear sigma model technology of \cite{Witten_phases}. Our techniques 
allow us to go further and perform a systematic analysis of all periods. 
In Section 5, we will use the results derived below 
in order to address certain puzzles
about the small radius limit of this model.

In this example,  $\psi$  
is related to the hypergeometric coordinate through $z=\psi^{-8}$. The 
associated hypergeometric symbol is 
{\footnotesize $\HG_symbol[{1/2,1/2,1/2,1/2},{1,1,1}]$}, so the model 
fits into the scheme discussed above for the particular value $\alpha=1/2$. 
The canonical and Jordan forms of the monodromy about $z=\infty$, as 
well as a choice for the matrix $P$ are given in Appendix A, 
while the Meijer monodromies are given by:

{\footnotesize
\be
T[0]=\left [\begin {array}{cccc} 1&0&0&0\\-2i
\pi &1&0&0\\-4{\pi }^{2}&-2i\pi &1&0
\\0&0&-2i\pi &1\end {array}\right ]
~~,~~
T[\infty]=
\left [\begin {array}{cccc} -7&-4{\frac {i}{\pi }}&4{\pi 
}^{-2}&2{\frac {i}{{\pi }^{3}}}\\-2
i\pi &1&0&0\\-4{\pi }^{2}&-2i
\pi &1&0\\0&0&-2i\pi &1\end {array}
\right]~~,
\ee}
and $T[1]=T[0]^{-1}T[\infty]$. These matrices satisfy:
\be
(T[0]-I)^4=0~~,~~(T[1]-I)^2=0~~,~~(T[\infty]^2-I)^4=0~~\nn~~.
\ee\noindent Note that the matrix $T[\infty]$ is {\em not} maximally unipotent.

Partial information about the integral structure is provided by a set of 
periods associated (up a {\em common factor}) with a basis of a full 
sublattice of $H_3(Y,\Z)$:
{\footnotesize
\be
U_E(z)=EU(z)~~,\mbox{with~}~E=
\left [\begin {array}{cccc} 1&0&0&0\\-7&-4{\frac {
i}{\pi }}&4{\pi }^{-2}&2{\frac {i}{{\pi }^{3}}}
\\25&16{\frac {i}{\pi }}&-20{\pi }^{-2}
&-12{\frac {i}{{\pi }^{3}}}\\-63&-44{
\frac {i}{\pi }}&56{\pi }^{-2}&38{\frac {i}{{
\pi }^{3}}}\end {array}\right ]~~.
\ee}
In this case, one obtains two 
weakly integral periods vanishing at $z=1$:
{\footnotesize
\bea
\label{vanishing}
U_{v1}&=&\frac{2i}{\pi^3}[U_3-2\pi^2 U_1]=[5,6,4,1] U_E~~,\\
U_{v2}&=&-\frac{8}{\pi^2}[U_2+i\pi U_1-2\pi^2 U_0]
=[15,11,5,1]U_E~~.\nn
\eea}\noindent In the mirror picture, these correspond to a $D6$ and a 
$D4$-brane which become massless at $z=1$. In fact, any linear combination of 
these periods will also vanish there, so we can for example also 
consider the vanishing period $U_{v1}+U_{v2}$, which in the mirror picture 
also corresponds to a collapsing $D6$-brane. This situation will be discussed 
in more detail in Section 5.

In this example, the constant $\kappa=e^{4(\psi(\alpha)-\psi(1))}$ has 
the value $2^{-8}=\frac{1}{256}$. 
Figure 5 displays the absolute values of the special coordinate $t$ and of 
the weakly integral periods $U_{v1},~U_{v2}$ as 
functions of the imaginary part 
$s=\frac{4}{\pi}\log\frac{2}{|z|}$ of the algebraic coordinate
on the moduli space. 
In Figure 6 we plot the absolute value of the special coordinate $t$ as a 
function of $s$, including the region $s<0$ of the moduli space, which has no 
classical analogue. In this example, we have $\psi(\alpha)=\psi(1-\alpha)=
\psi(1/2)$, so the asymptotic form of $t$ for $|z|\rightarrow \infty$ is:
\be
t_{as}=-\frac{1}{2}+\frac{3i\pi}{2\log z}~~.
\ee\noindent In particular, 
$J={\rm Im}(t)\approx \frac{3}{2}\frac{\pi}{\log |z|}$ 
remains nonnegative for $|z|>>1$, as pointed out
\footnote{The reader should note that the variable $z$ used in equation 
(37) of \cite{geometric_interpretation} is the {\em inverse} of the 
hypergeometric coordinate $z$ used in the present paper.}
in \cite{geometric_interpretation}.

\vskip 0.5 in
\begin{center}
$\begin{array}{cc}
\begin{array}{c}\scalebox{0.3}{\begin{picture}(0,0)%
\epsfbox{7_2222.pstex}%
\end{picture}%
\setlength{\unitlength}{3947sp}%
\begingroup\makeatletter\ifx\SetFigFont\undefined%
\gdef\SetFigFont#1#2#3#4#5{%
  \reset@font\fontsize{#1}{#2pt}%
  \fontfamily{#3}\fontseries{#4}\fontshape{#5}%
  \selectfont}%
\fi\endgroup%
\begin{picture}(10828,7845)(1168,-8173)
\put(1426,-961){\makebox(0,0)[lb]{\smash{\SetFigFont{20}{24.0}{\rmdefault}{\bfdefault}{\updefault}$m$}}}
\put(4876,-7561){\makebox(0,0)[lb]{\smash{\SetFigFont{20}{24.0}{\rmdefault}{\bfdefault}{\updefault}$s$}}}
\end{picture}
}\end{array}&
\begin{array}{c}\scalebox{0.39}{\begin{picture}(0,0)%
\epsfbox{7_2222t.pstex}%
\end{picture}%
\setlength{\unitlength}{3947sp}%
\begingroup\makeatletter\ifx\SetFigFont\undefined%
\gdef\SetFigFont#1#2#3#4#5{%
  \reset@font\fontsize{#1}{#2pt}%
  \fontfamily{#3}\fontseries{#4}\fontshape{#5}%
  \selectfont}%
\fi\endgroup%
\begin{picture}(9628,6745)(751,-7073)
\put(751,-3361){\makebox(0,0)[lb]{\smash{\SetFigFont{17}{20.4}{\rmdefault}{\bfdefault}{\updefault}$|t|$}}}
\end{picture}
}\end{array}\\
\begin{array}{c}
~\\
\mbox{Figure 5. {\footnotesize  
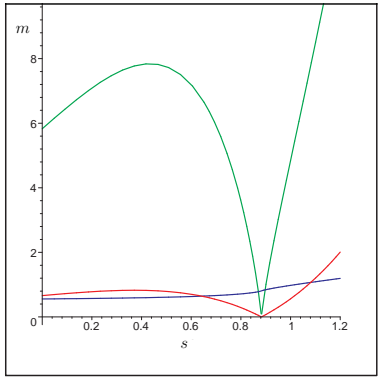
Graph of $|U_{v1}|,|U_{v2}|$ and $|t|$ versus the imaginary}}\\ 
\mbox{{\footnotesize part $s$ of the algebraic coordinate for 
$s \in [0,~1.2]$. The }}\\ 
\mbox{{\footnotesize point $z=1$ corresponds to $s=\frac{4{\rm log} 2}{\pi}
\approx 0.882$.}}\\
\end{array}
&
\begin{array}{c}
~~\\
~~\\
\mbox{Figure 6. {\footnotesize  
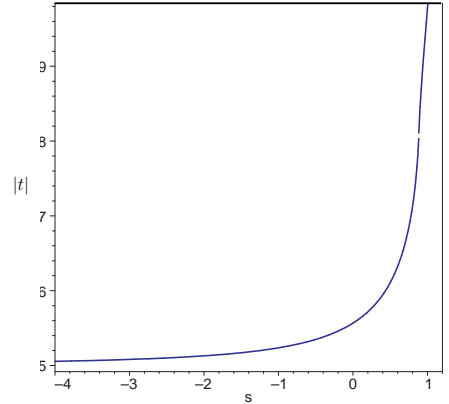
Graph of $|t|$ versus $s$}}\\
\mbox{{\footnotesize for $s \in [-4,1.2]$.}}\\
~~
\end{array}~~~~~~~~~~
\end{array}$~~~~~~~~~~~~~~~~~~~~~~~~~~~~~~~~~~~~~~~~~~~~~~~~~~~~~~~~~~~~~~~~~~~~~~~~~~~~~~~~~~~~~~~~~~~~~~~~~~~~~~~~~~~~~~~~~~~~~~~~~~~~~~~~~~~~~~~~~~~~~~~~~~~~~~~~
\end{center}

\section{Small/large radius duality}

\subsection{Basic considerations}

The results we have obtained for the model $\P^7[2,2,2,2]$ 
apparently preclude us from interpreting $z=\infty$ 
as a large complex structure point. 
Indeed, the associated monodromy matrix is not maximally unipotent, 
but rather satisfies $(T[\infty]^2-I)^4=0$. This behaviour 
is due to the factors of 
$z^{-\alpha}=z^{-1/2}$ in the 
expansions (\ref{LG_expansions4}) of the periods for $|z|>1$. One may be 
tempted to interpret this result as showing that the limit $z\rightarrow 
\infty$ of the model does not admit a standard geometric (i.e. Calabi-Yau) 
description \cite{geometric_interpretation}. However, the form of 
the monodromy about $z=\infty$ is tantalizing close to that of the monodromy 
about a large complex structure point, which is an indication that something 
more interesting may be going on. 

Indeed, it was noticed in \cite{Candelas_periods} that the 
moduli space of this model admits a symmetry $z\rightarrow 1/z$. This follows 
by replacing $\psi$ with its inverse and performing the 
change of coordinates:
\bea
\label{change_coords}
\begin{array}{ccccc}
x_1&=&y_1+iy_2~~,~~x_2&=&iy_1+y_2\\
x_3&=&y_7+iy_8~~,~~x_4&=&iy_7+y_8\\
x_5&=&y_5+iy_6~~,~~x_6&=&iy_5+y_6\\
x_7&=&y_3+iy_4~~,~~x_8&=&iy_3+y_4
\end{array}~~,
\eea\noindent which preserves the form of the defining equations (\ref{Y}).
Hence the manifolds $Y_\psi$ and $Y_{\frac{1}{\psi}}$ described
by (\ref{Y}) for the parameters $\psi$ and $\frac{1}{\psi}$ are isomorphic, 
which implies that the nature of the points $z=0$ and 
$z=\infty$ is identical. Indeed, the isomorphism between 
$Y_\psi$ and $Y_{\frac{1}{\psi}}$ forces us to conclude
that the points $z=0$ and $z=\infty$ are physically indistinguishable --- 
this is an {\em exact} statement in the full IIB 
string theory on $Y$, since its vector multiplet moduli space 
does not receive quantum corrections \cite{strominger_vm}.
This, however, seems to be at odds with the different behaviour of the Meijer 
periods in the two limits $z=0$ and $z=\infty$.

In order to clarify the 
situation, let us consider the effect of the change of 
variable $z\rightarrow s:=1/z$ on the hypergeometric equation (\ref{HG_eq2}).
Under this operation, the equation is transformed into:
\be
\label{HG'}
\left[s\delta'^4-(\delta'-\alpha)^4\right]{\tilde u}(s)=0~~,
\ee\noindent 
where $\delta'=s\frac{d}{ds}=-z\frac{d}{dz}$ and 
${\tilde u}(s):=u(1/s)$. Thus (\ref{HG_eq2}) is {\em not} invariant 
under this symmetry. However, it is not hard to see that the form of 
(\ref{HG_eq2}) 
is preserved under the {\em combined} change of variable and function:
\bea
\label{impl}
&z&\rightarrow s:=\frac{1}{z}~~\\
&u&\rightarrow u':=z^{-\alpha}u~~,\nn
\eea\noindent i.e. $u(z)\rightarrow z^{-\alpha}u(1/z)$. Since 
$u(z)=\int_{\gamma}{\Omega(z)}$ is the period of the holomorphic 3-form 
$\Omega$ on a 3-cycle $\gamma\in H_3(Y,\Z)$, it follows that the 
implementation of the symmetry $z\rightarrow \frac{1}{z}$ requires 
a rescaling of $\Omega$:
\be
\Omega(z)\rightarrow z^{-\alpha}\Omega(1/z)~~.
\ee\noindent 

What, then, is the correct interpretation of the point $z=\infty$ ?
The answer follows by recalling that 
the moduli space of the closed conformal field theory on 
$Y$ is built by considering marginal deformations, a process which is  
analytic in the deformation parameter $z$. This forces the periods to 
have different 
behavior in the regions $|z|<1$ and $|z|>1$. 
There is, however, 
a basic point to take into account: when performing marginal deformations 
one must specify a starting point !
In fact, one could as well choose this point to be $z=\infty$ and use 
the periods ${\tilde U}_j(z)=U_j(1/z)$ instead of $U_j(z)$.  Therefore,  
the interpretation of $z=0$ and $z=\infty$ as ``large'' and 
``small'' radius points is indeed conventional and can be reversed, 
even though the analytic continuations of the associated periods do not 
coincide. In fact, interchanging these points corresponds to 
starting on different branches of a double cover of the moduli space.
This follows by noticing that, since $Y_z$ and 
$Y_{\frac{1}{z}}$ are isomorphic, the complex structure moduli space of 
$Y$ is not the copy of $\P^1$ parameterized by $z$, but rather its quotient 
${\cal M}$ via this identification. This quotient is again a $\P^1$, which 
can be parameterized, for example, by the variable:
\be
x=\frac{2z}{z^2+1}~~.
\ee
The map $z\rightarrow u$ gives a double cover of ${\cal M}$, 
branched over the points $x=+1$ and $x=-1$, 
which are the images of 
$z=1$ and $z=-1$, respectively. The unit circle 
$|z|=1$ is mapped 
into the region $x \in [-\infty,-1]\cup [1,\infty]$, which represents 
a segment on the associated Riemann sphere (see Figure 7). The points $z=0$ 
and $z=\infty$ are both mapped into the point $x=0$. Since these points lie 
on different branches of our double cover, picking one of them as the 
large complex structure limit amounts to choosing a particular 
realization of the model. 

\vskip 0.5in 
\hskip 1.7in\scalebox{0.5}{\input{cover.pstex_t}}
\vskip 0.3in
\hskip 1in \begin{center}Figure 7. {\footnotesize  
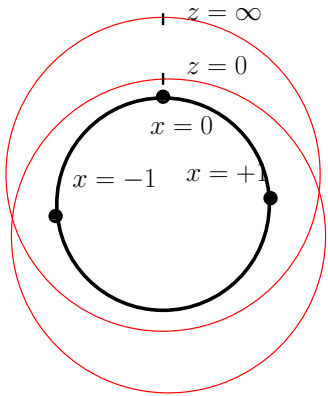
The coordinate $z$ parameterizes a double cover of the complex structure 
moduli space of $Y$. The figure shows the topology of the restriction of this 
cover above the circle ${\rm Im}(x)=0$ on the Riemann sphere of $x$.} 
\end{center}

This situation is similar to the standard interpretation of T-duality 
for the conformal field theory on a circle. 
In that case, marginal deformations starting from a point $R>R_0$
build the continuation of the theory 
through the self-dual point $R_0=\sqrt{\alpha'}$, into the region 
$R<R_0$. The duality $R\approx {\tilde R}=\alpha'/R$ 
identifies this continued 
theory with its form at a radius ${\tilde R}>R_0$, but this discrete 
identification is not captured by the marginal deformations. Just as in the 
case of T-duality, the global 
identification $z\approx 1/z$ in our model is ``accidental'' in the 
sense that it is not captured by marginal deformations associated with the 
$(c,c)$ ring.

In order to make this more precise, let us compute the action of our 
symmetry on $H^3(Y,\C)$. Consider acting with the transformation
(\ref{impl}) on the Meijer periods:
\be
\label{M_tf}
U_j(z)\rightarrow U'_j(z)=z^{-1/2}U_j(1/z)~~.
\ee\noindent Since (\ref{impl}) is a symmetry of (\ref{HG_eq2}), it follows 
that both $(U_j)_{j=0..3}$ and $(U'_j)_{j=0..3}$ give a basis of solutions. 
Hence  
there must exist a constant matrix $C=(c_{ij})_{i,j=0..3}$ such that 
$U'_i(z)=c_{ij}U_j(z)$. In fact, this conclusion is a bit too quick, 
since the functions $U_j$ are multi-valued, so we must be careful to take 
the branch-cuts into account. 
The correct statement is that such a relation must hold 
on every open and connected subset $V$ of the moduli space which does 
not intersect the cuts. In fact, the matrix $C$ can depend on $V$
\footnote{In mathematical parlance, 
$C$ is a locally constant matrix -valued function 
defined on the moduli space with the branch-cuts removed.}. 
Since the set $V_0=\{|z|<1\}$ does not contain any cuts (see Figure 2), 
it suffices to 
start by considering our relation in this region. 
Hence we define $C$ to be the 
matrix associated with $V_0$. Then performing a transformation 
$z\rightarrow 1/z$ shows that the matrix associated with the region 
$V_1=\{|z|>1\}$ is the inverse of $C$. Thus, we expect the relations:
\bea
\label{rel}
z^{-1/2}U(1/z)&=&CU(z)~~\mbox{,~~~~if~}~|z|<1~~\\
z^{-1/2}U(1/z)&=&C^{-1}U(z)~~\mbox{,~if~}~|z|>1~~.\nn
\eea

In order to check these equalities and determine the matrix $C$, 
let us take $z$ to be such that $|z|<1$.
Then $|\frac{1}{z}|>1$ and we have:
\bea
U_j(z)&=&\frac{(-1)^j}{j!}\sum_{n=0}^{\infty}{
\left[\frac{(\frac{1}{2})_n}{n!}\right]^4\nu_j(n,z)z^n}~~\\
U'_i(z)&=&\frac{(\delta_{i,odd}-i\delta_{i,even})}{6\pi^{3-i}}
\sum_{n=0}^{\infty}{\left[\frac{(\frac{1}{2})_n}{n!}\right]^4
\mu_i(n,1/z)z^n}~~.
\eea \noindent Defining $b_{ij}$ via:
\be
c_{ij}=(-1)^j j!
\frac{(\delta_{i,odd}-i\delta_{i,even})}{6\pi^{3-i}}b_{ij}~~,
\ee\noindent it suffices to compute the matrix $B=(b_{ij})_{i,j=0..3}$, which 
satisfies:
\be
\mu_i(n,1/z)=b_{ij}\nu_j(n,z)~~.
\ee
Using the explicit form of these sequences given in 
(\ref{nus},\ref{etas}) and (\ref{mus},\ref{xis}), 
it is not very hard to show that the required matrix has the form:
{\footnotesize
\be
B=
\left [\begin {array}{cccc} -6\,i{\pi }^{3}&18\,{\pi }^{2}&3
\,i\pi &-1\\0&6\,{\pi }^{2}&0&-1
\\0&12\,{\pi }^{2}&3\,i\pi &-1
\\0&0&0&-1\end {array}\right ]~~,
\ee}\noindent which finally leads to the transition matrix of interest:
{\footnotesize
\be
\label{C}
C=\left [\begin {array}{cccc} -1&3{\frac {i}{\pi }}&{\pi }^{-
2}&-{\frac {i}{{\pi }^{3}}}\\0&-1&0&{\pi}^
{-2}\\0&2i\pi &1&-{\frac {i}{
\pi }}\\0&0&0&1\end {array}\right ]\Rightarrow
C^{-1}=\left [\begin {array}{cccc} 
-1&-{\frac {i}{\pi }}&{\pi }^{-2}&{\frac {i}{{\pi }^{3}}}\\0&-1&0&{\pi }^{-2
}\\0&2i\pi &1&-{\frac {i}{\pi }}
\\0&0&0&1\end {array}\right ]~~.
\ee}

The geometric interpretation  of these results follows by 
writing the Meijer periods in the form:
\be
U_j(z)=\int_{g_j(z)}{\Omega(z)}~~,
\ee\noindent where $g_j(z)~(j=0..3)$ is a basis of $H_3(Y_z,\C)$. 
Following the 
general theory of variations of Hodge structure (see \cite{morrison_aspects} 
for a review in the context of its applications to mirror symmetry), we  
take the classes $g_j$ to be flat with 
respect to the Gauss-Manin connection on the moduli space
\footnote{Usually one takes this connection to act on cohomology, 
but here we use Poincare duality to transport the local system from 
$H^3(Y)$ to $H_3(Y)$.
Hence we think of $g_i(z)$ as being flat sections of a bundle with fiber 
$H_3(Y_z)$. Then $g_i(z)$ will be multivalued due to the nontrivial holonomy 
of the connection.}.
Here $\Omega$ is normalized such that $\lim_{z\rightarrow 0} U_0(z)=1$.
On the other hand, using the variable $s=1/z$ and starting with $s=0
\iff z=\infty$ as the large complex structure point (in which case the 
Picard Fuchs equation coincides with the equation obtained from 
(\ref{HG_eq2}) by substituting $s$ for $z$) gives 
Meijer periods ${\tilde U}_j(s)=U_j(1/s)$, which can also be written in 
the form:
\be
{\tilde U}_j(s)=\int_{{\tilde g}_j(s)}{{\tilde \Omega}(s)}~~,
\ee\noindent where ${\tilde g}_j(s)$ is a flat basis of $H_3(Y,\C)$ while 
${\tilde \Omega}$ is the holomorphic 3-form on $Y$ normalized 
via $\lim_{s\rightarrow 0} {\tilde U}_0(s)=1$. Then (\ref{rel})
shows that:
\bea
{\tilde g}_i(z)&\equiv &C_{ij}g_j(z)~~\\
{\tilde \Omega}(z)&\equiv& z^{1/2}\Omega(z)~~,
\eea\noindent for $|z|<1$. 
It follows that $C$ encodes the relation between the Meijer bases 
$g_i$ and ${\tilde g}_i$ of $H_3(Y,\C)$ 
associated with the points $z=0$ and $z=\infty$, while the 
rescaling by $z^{1/2}$ reflects the different normalization of $\Omega$ 
required by their interpretation as large complex structure points.

We can now shed more light on the vanishing periods 
at $z=1$. Indeed, applying (\ref{rel}) at that point shows that 
the vector 
{\footnotesize $U(1)=\left[\begin{array}{c}U_0(1)\\U_1(1)\\U_2(1)\\U_3(1)
\end{array}\right]$} is an eigenvector of $C$ with eigenvalue one:
\be
(C-I)U(1)=0~~.
\ee\noindent The kernel of the matrix $(C-I)$ is a two-dimensional subspace
spanned by the row vectors
\footnote{The matrix $C$ has eigenvalues $-1$ and $+1$, each of which have 
multiplicity two. However, it is easy to check that $C$ is not diagonalizable.
The reader may wonder why we do not apply relation (\ref{rel}) to the other 
fixed point $z=-1$ and try to obtain vanishing periods there via a 
similar argument. The reason is, of course, that $-1$ is a branch point for 
our 
analytic continuations, so that the limit of $U(z)$ at this point is not 
well-defined. While (\ref{rel}) holds in a directional 
limiting sense at $z=-1$ 
(no matter from what direction in the complement of the cut we approach 
that point), this does not imply vanishing of a period there since the 
limits of $U(z)$ and $U(1/z)$ are different as $z$ approaches the value $-1$
(note that $z$ and $1/z$ lie on different sides of the cut).}:
\bea
\left[0, -2\pi^2, 0, 1 \right]~~,~~
\left[-2\pi^2, i\pi, 1, 0 \right]~~\nn
\eea\noindent associated with the vanishing periods (\ref{vanishing}). 
This reproduces the result of Section 4 that this model admits 
a two dimensional subspace of periods which vanish at $z=1$. It also shows 
that this somewhat unusual situation is a consequence of the 
``accidental'' symmetry $z\rightarrow 1/z$.

\subsection{Physical interpretation}

\subsubsection{The closed string sector}

Let us consider the implications of these results for the bulk conformal 
field theory associated with our compactification.  The B-model  
defined by $Y$ contains chiral 
primary operators ${\cal O}^{p,p}$ which are in one to one 
correspondence with generators of the Hodge groups $H^{p,3-p}(Y)$.
When computing correlators, we can replace $H^{p,3-p}(Y)$ with their 
holomorphic counterparts ${\cal H}^{p,3-p}={\cal F}^{3-p}\cap {\cal W}_p$, 
where:
\be
0\subset {\cal F}^0\subset {\cal F}^1\subset {\cal F}^2\subset {\cal F}^3=
H^3(X)~~
\ee\noindent is the Hodge filtration and:
\be
0\subset 
{\cal W}_0\subset {\cal W}_1\subset {\cal W}_2\subset {\cal W}_3=H^3(Y)~~
\ee\noindent is the ``reduced'' monodromy weight filtration associated with a 
large complex structure point (see Appendix A of \cite{branes} for a short 
explanation of this concept). Roughly, ${\cal W}_j$ is the space of those 
periods which have $\log^j$ 
leading behaviour near that point
\footnote{The vector space ${\cal H}^3(Y_z)$ can be 
identified with the space spanned by the vectors {\scriptsize 
$w_j:=\left[\begin{array}{c}
U_j(z)\\\delta U_j(z)\\\delta^2 U_j(z)\\\delta^3 U_j(z)\end{array}\right]$}. 
Viewing $w_j$ as a set of initial conditions for the Picard Fuchs equation at 
the point $z$ further identifies this space with the space of solutions to  
(\ref{HG_eq2}). }. 
The monodromy filtrations can be easily determined by making use of the 
special logarithmic behaviour of the Meijer periods (see the 
expansions (\ref{LCS_expansions})): 
if $z=0$ is treated as a large complex structure point, then we obtain 
a filtration ${\cal W}$ which can be identified with the spaces of 
periods spanned 
by:
\be
{\cal W}_0=<U_0>~,~{\cal W}_1=<U_0,~U_1>~,~{\cal W}_2=<U_0,~U_1,~U_2>~,~
{\cal W}_3=<U_0,~U_1,~U_2,~U_3>~~.
\ee\noindent On the other hand, treating $z=\infty$ as a large complex 
structure point gives:
\be
{\tilde {\cal W}}_0=<{\tilde U}_0>~,~{\tilde {\cal W}}_1=<{\tilde U}_0,~
{\tilde U}_1>~,~
{\tilde {\cal W}}_2=<{\tilde U}_0,~{\tilde U}_1,~
{\tilde U}_2>~,~{\tilde {\cal W}_3}=<{\tilde U}_0,~{\tilde U}_1,
{\tilde U}_2,~{\tilde U}_3>~~.
\ee Hence (\ref{rel}) implies a nontrivial relation between 
$({\tilde {\cal W}})$ and 
$({\cal W})$ and thus between ${\tilde {\cal H}}^{p,3-p}(Y_z)$ and 
${\cal H}^{p,3-p}(Y_z)$. It follows that our symmetry involves a  
``rotation'' of the chiral primary operators ${\cal O}^{p,3-p}$.

\subsubsection{The D-brane sector}

The (BPS saturated) 
D-brane sector of our compactification can be realized by considering 
the open conformal field theory or, equivalently, by including boundary states.
In the large complex structure limit of the 
IIB theory on $Y$, these correspond to special Lagrangian cycles 
$C$ in $Y$. 
Hence given a boundary state we can associate to it the homology class 
$\gamma=[C]$ 
of the associated cycle and hence the corresponding period 
$\int_{\gamma}{\Omega}$. As we move away from this limit, 
the correspondence may be destroyed for some boundary states, due 
to the fact that the path we use for performing the marginal 
deformations could cross a marginal stability line \cite{Douglas_quintic}. 
On such a line, the associated special Lagrangian cycle is expected to suffer 
a splitting transition of the type discussed in \cite{Joyce,Kachru_Greevy}. 
Since we do not have a proper understanding of marginal stability lines in this
model, the conclusions we can derive regarding the behaviour of D-brane states 
are only tentative. 

The most basic question about such states concerns the dimensionality of the 
type IIA D-brane on $X$ mirror to a given IIB D-brane on $Y$. As discussed 
in \cite{Ooguri,quantum_volumes,Morrison_II}, this is determined by the 
order of the logarithmic behaviour of the associated period in the large 
complex structure limit, i.e. by the smallest component of the monodromy 
weight filtration which contains that period. In our model, we have {\em two}
points which can play the role of large complex structure points, and hence 
two monodromy weight filtrations $({\cal W})$ and $({\tilde {\cal W}})$. 
Thus the 
correspondence between the mirror D-brane states (and even their dimension) 
involves a nontrivial rotation of $H_{even}(X)$.  

A rather dramatic effect of this type can be observed as follows.
Suppose that we define the large radius/large complex structure 
limit to correspond to $z=0$. 
Then consider a $D2$-brane in the large 
radius limit on $X$, whose 
mirror D3-brane is associated (up to a factor) with the period 
$U_1$. Note that
this period  is weakly integral (i.e. proportional with the 
period of $\Omega$ over an {\em integral} homology class of $Y$).
Now perform marginal deformations until we cross the circle $|z|=1$, 
reaching a point $z_0$ which lies outside the unit disk. At this 
point, we have a boundary state (the deformation of the original D-brane state)
in the conformal field theory associated with $z_0$. Performing a duality 
transformation maps this theory into an equivalent conformal theory 
for which the large radius point correspond to $z=\infty$; this transformation 
will modify the associated period through the action of $C^{-1}$ 
(and rescaling by $z_0^{1/2}$). Inspection of the matrix $C^{-1}$ 
(equation (\ref{C})) shows 
that the associated period has $\log^2$ behaviour around $z=\infty$, and 
hence the mirror boundary state corresponds to a D4-brane! In fact, choosing 
$z_0$ to be far away in the $z$-plane assures that we are in the 
large radius limit of the dual model, and hence the associated $D4$-brane 
must correspond to a holomorphic 
4-cycle on $X$. In other words, our duality seems to 
identify some $D2$-brane states on $X$ with $D4$-branes.  
Of course, this surprising 
conclusion may 
be avoided if the path used for analytic continuation 
crosses a marginal stability curve, or if the homology class under 
consideration does not actually contain a special Lagrangian cycle.

\subsection{Small versus large size}

What is the action of our symmetry on the size of $X$ ? The answer to this 
question depends  
on the precise definition of ``size''. Let us first consider 
the nonlinear sigma model measure of \cite{small_distances1}, which was 
shortly reviewed in the introduction. Following \cite{small_distances1}, 
we can start with $z=0$ as the large radius 
point and measure size by using the analytic continuation of the special 
coordinate $t(z)$. Then the symmetry $z\rightarrow 1/z$ identifies $t(z)$ 
with ${\tilde t}(z):=t(1/z)$. Eliminating $z$ defines a map ${\tilde t}=f(t)$, 
which can be determined 
numerically and is plotted in Figure 8 for $|z|$ belonging to the interval 
$(1,~10^4)$ (${\tilde J}$ remains positive in this range, even though this 
is not obvious at the scale and from the viewing angle of this figure). 
We see that the duality indeed maps small into large 
distances --- a conclusion which is now established at the {\em quantum}
level. Figure 9 displays the values of $t(z)$ for $|z|\in (1,~10^8)$. 
Note that 
$J={\rm Im}(t)$ remains positive when ${\rm Im}(z)\neq 0$. 

\vskip 0.5 in
\begin{center}
$\begin{array}{cc}
\begin{array}{c}\scalebox{0.35}{\input{dualgraph.pstex_t}}\end{array}&
\begin{array}{c}\scalebox{0.35}{\begin{picture}(0,0)%
\epsfbox{timage.pstex}%
\end{picture}%
\setlength{\unitlength}{3947sp}%
\begingroup\makeatletter\ifx\SetFigFont\undefined%
\gdef\SetFigFont#1#2#3#4#5{%
  \reset@font\fontsize{#1}{#2pt}%
  \fontfamily{#3}\fontseries{#4}\fontshape{#5}%
  \selectfont}%
\fi\endgroup%
\begin{picture}(10828,7845)(1093,-7648)
\put(4576,-7186){\makebox(0,0)[lb]{\smash{\SetFigFont{20}{24.0}{\rmdefault}{\bfdefault}{\updefault}$B$}}}
\put(1351,-3511){\makebox(0,0)[lb]{\smash{\SetFigFont{20}{24.0}{\rmdefault}{\bfdefault}{\updefault}$J$}}}
\end{picture}
}\end{array}\\
\begin{array}{c}
~\\
\mbox{Figure 8. {\footnotesize  
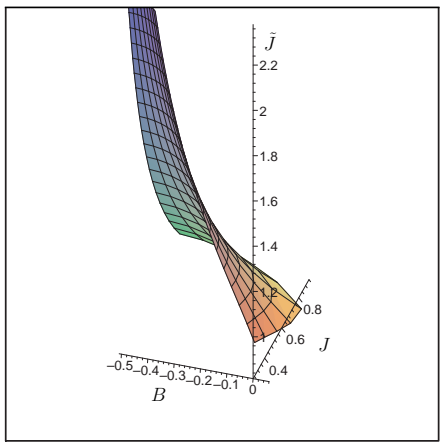
Graph of ${\tilde J}=Im({\tilde t})$ vs $t=B+iJ$.~~~~~~~~~~~~~~~~~~~}}  
\end{array}
&
\begin{array}{c}
~~\\
~~\\
\mbox{Figure 9. {Values of $t(z)$ for $1<|z|<10^8$.~~~~~~}}\\
~~
\end{array}~~~~~~~~
\end{array}$~~~~~~~~~~~~~~~~~~~~~~~~~~~~~~~~~~~~~~~~~~~~~~~~~~~~~~~~~~~~~~~~~~~~~~~~~~~~~~~~~~~~~~~~~~~~~~~~~~~~~~~~~~~~~~~~~~~~~~~~~~~~~~~~~~~~~~~~~~~~~~~~~~~~~~~
\end{center}

What about the quantum volume of $X$ ? As discussed in the introduction, 
this is measured by the mass of a D6-brane wrapped over $X$, and it is 
natural to pick the D6-brane state whose mass vanishes at $z=1$, which is 
plotted in Figure 5. There is no 
positive lower bound for the (quantum) volume of $X$ --- 
string theory allows the entire manifold to shrink to zero size.

\subsection{Phases}

The semiclassical Kahler moduli space of $X$ 
was studied in \cite{geometric_interpretation}, where it was shown that 
the model admits two phases, one of which is a large radius Calabi-Yau phase. 
The other phase can be analyzed via the linear sigma 
model techniques of \cite{Witten_phases}, with the result that it is a hybrid 
phase which can be roughly described as a fibration of a $\Z_2$ 
Landau-Ginzburg orbifold over a $\P^3$. This picture is tantalizingly close 
to a purely geometric description of that phase (say, in terms of a 
nonlinear sigma model having $\P^3$ or a closely related space as a target, 
for example through a construction along the lines of 
\cite{non_CY1,non_CY2,non_CY3}) 
but, as pointed out in \cite{geometric_interpretation}, the semiclassical 
picture provided by the 
linear sigma model is affected by strong quantum corrections which have the 
potential to seriously modify the discussion, thus making this geometric 
interpretation inconclusive. Our results allow us to make a precise statement 
about the effect of these corrections: they modify the theory in such a way 
that it becomes equivalent with its large radius incarnation ! In fact, 
once quantum corrections have been taken into account, there is no physical 
difference between the two limits and the model has a 
single phase (see Figure 10).

\

\hskip 1.2in\scalebox{0.4}{\input{phases.pstex_t}}

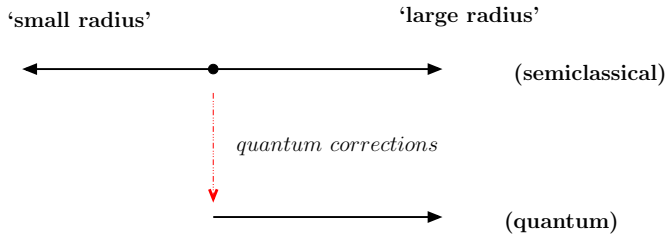
\begin{center}Figure 10. {\footnotesize The effect of quantum corrections
on the phase diagram of the IIA compactification on $X$.}
\end{center}

\subsection{Interpretation via special Lagrangian fibrations}

How can we understand the behaviour of this model from the point of 
view of the SYZ conjecture \cite{SYZ} ?. Since both  
$z=0$ and $z=\infty$ can be viewed as large complex structure points, 
the natural expectation is that $Y_z$ should admit two special 
Lagrangian fibrations, well-defined on some vicinities of the 
points $z=0$ and $z=\infty$, and related by the transformation 
(\ref{change_coords}). 
It was shown in \cite{Gross1,Gross3} 
that the monodromy weight filtration is determined by the fibration. 
Hence using one or the other of these fibrations corresponds to declaring 
$z=0$ or $z=\infty$ to be the large complex structure point. 
Then our small-large radius duality 
appears as a consequence of the fact that the two fibrations
are isomorphic.

The techniques for constructing 
special Lagrangian fibrations of Calabi-Yau manifolds are not yet fully 
developed (see \cite{Ruan1, Ruan2, Zharkov, Gross1, Gross2, Gross3, Hitchin} 
for partial results in this direction), so it is premature to attempt a 
complete analysis along these lines. 
However, simple and powerful methods are available 
in the large complex structure limit \cite{SYZ, Zharkov, Ruan1, Gross3, LV}, 
where the problem can be reduced to one of toric geometry and hence can be 
approached with the machinery available in such situations \cite{toric}. 
Appendix B uses a simple generalization of these techniques in order to
identify the {\em topology} of the relevant fibrations. 
As in the hypersurface case, the base of each fibration turns out 
to be a 3-sphere.

The SYZ picture provides a natural interpretation of the nontrivial action 
of the duality on D-brane states: since mirror symmetry amounts to T-duality 
along the $T^3$ fibers, the dimension of the mirror D-brane depends on the 
relative 
position of a given  IIB D3-brane with respect to the fibration of interest. 
Changing the fibration modifies this relative position, and hence can modify 
the dimension of the mirror holomorphic cycle. This is just the  
familiar fact that the dimension of a D-brane increases or decreases 
when performing T-duality along a direction orthogonal or parallel 
with its  volume.

\section{Conclusions}

We completed the study of the hierarchy of one-parameter models introduced 
in \cite{branes}, providing more evidence that the phenomena discussed in 
that paper are generic: in a typical IIA compactification on a 
one-parameter Calabi-Yau manifold, the non-perturbative state which becomes 
massless at the mirror of the conifold point is associated with a 
D6-brane. The general results derived in 
\cite{branes} and in the present paper should open the way for extensions 
of the work of \cite{Douglas_quintic} to more general Calabi-Yau 
compactifications,
as well as providing a convenient framework for a systematic study of 
issues of marginal stability (see \cite{Douglas_quintic,Joyce,Kachru_Greevy} 
for a few steps in this direction) through the effective field theory methods 
of \cite{Moore_arithmetics}.

From a methodological point of view, our results show that most one-parameter
models fit into a hypergeometric hierarchy, which allows for a very systematic
approach to the computation of all periods. This should help prepare 
the ground for 
further investigations of D-brane effects in Calabi-Yau 
compactifications. The universal large radius 
expansions we have obtained should also 
help clarify some of the arithmetic properties 
of the mirror map when combined with the work of \cite{arithmetic_mirror1} 
and  \cite{arithmetic_mirror2}.

We also performed a detailed study of a special one-parameter example, which 
displays some unusual features. In particular, we were able to bring some 
detailed evidence that this model realizes a Calabi-Yau version of 
large-small radius 
duality, thus confirming the suspicions of 
\cite{Candelas_periods, geometric_interpretation}. We also presented evidence 
that, in the framework of 
\cite{SYZ}, this duality is realized through the existence of {\em two}
special Lagrangian fibrations --- a feature which has 
interesting implications for the physics of D-branes in the associated 
string theory compactification. It would be interesting to 
investigate this phenomenon further, as well as its implications for the problem of 
marginal stability of D-brane states. 
Since the duality exchanges the small and radius points, it should 
be possible to use it in this model in 
order to extract strong results regarding this issue. 

Another interesting question is to what extent these phenomena 
generalize. Multiple large complex structure points 
are common in multi-parameter models (any model admitting 
topology-changing transitions possesses at least two such points). It would 
be interesting to see if similar discrete 
identifications occur in such models, and what can be learned from this 
about quantum corrections to the Kahler moduli space.

\appendix

\section{Some intermediate results for the models $\P^5[3,3]$ and 
$\P^7[2,2,2,2]$}

\subsection{The model $\P^5[3,3]$}
The canonical and Jordan form of the matrix $R[\infty]$, as well as a choice 
for the matrix $P$ are given below:
{\scriptsize
\bea
\begin{array}{c}
R_{can}[\infty]=
\left [\begin {array}{cccc} 0&-1&0&0\\0&0&-1&0
\\0&0&0&-1\\{\frac {4}{81}}&-4/9&{
\frac {13}{9}}&-2\end {array}\right ]~~,~~
R_J[\infty]=\left [\begin {array}{cccc} -2/3&1&0&0\\
0&-2/3&0&0\\0&0&-1/3&1\\0&0&0&-1/3
\end {array}\right ]~~\\
P=\left [\begin {array}{cccc} 2/3&5&4/3&-4\\4/9&8/3&4/
9&-8/3\\{\frac {8}{27}}&4/3&{\frac {4}{27}}&-4/3
\\{\frac {16}{81}}&{\frac {16}{27}}&{\frac {4}{81}}&
-{\frac {16}{27}}\end {array}\right ]
\end{array}~~\nn
\eea}\noindent 
The small radius 
arithmetic identity associated to the collapsing period at $z=1$ is:
$\sum_{n=0}^{\infty}{\frac{a_n}{n!^2}}=0$, where
{\footnotesize
\bea
a_n=\Gamma (n+1/3)^4\Gamma (-n+1/3)^2
\psi(-n+1/3)+\Gamma (n+1/3)^4\Gamma (-n+1/3)^2\psi(n+1)+~~~~~~~~~~~~~~~\nn\\
\Gamma (n+2/3)^4\Gamma(-n-1/3)^2\psi(-n-1/3)+
\Gamma (n+2/3)^4\Gamma(-n-1/3)^2\psi(n+1)-~~~~~~~~~~~~~~~\nn\\
2\Gamma (n+1/3)^4\Gamma (-n+1/3)^2\psi(-n+2/3)-\
2\Gamma (n+2/3)^4\Gamma (-n-1/3)^2\psi(-n+1/3)~~.~~~~~~\nn
\eea}\noindent A pair identity follows from the large radius 
expansions.

\subsection{The model $\P^7[2,2,2,2]$}
In this case, we have:
{\scriptsize 
\bea
\begin{array}{c}
R_{can}[\infty]=
\left [\begin {array}{cccc} 0&-1&0&0\\0&0&-1&0
\\0&0&0&-1\\1/16&-1/2&3/2&-2
\end {array}\right ]
~~,~~
R_J[\infty]=
\left [\begin {array}{cccc} -1/2&1&0&0\\0&-1/2&1&0\\0&0&-1/2&1\\0&0&0&-1/2
\end {array}\right ]\\
P=\left [\begin {array}{cccc} 1/8&1/4&1/2&1\\1/16&0&0&0
\\1/32&-1/16&0&0\\{\frac {1}{64}}&
-1/16&1/16&0\end {array}\right ]
\end{array}~~\nn
\eea}

\section{Special Lagrangian fibrations of $Y$}

A topological 
$T^3$ fibration in the large complex structure limit can be 
obtained by the methods of \cite{Zharkov}. This fibration is believed 
\cite{SYZ, Ruan1, Gross3,Zharkov} to 
admit a deformation to a special Lagrangian fibration of $Y$ as we move away 
from the large complex structure point. 
While the arguments discussed in those papers are restricted to hypersurfaces 
in toric varieties, our model $\P^7[2,2,2,2]$ is more general since it is a 
complete intersection. Assuming that some generalization of those 
arguments goes through in our case, we can attempt to construct 
our fibration along the same lines. 

For this, let us first consider the point $z=0\iff \psi= \infty$. 
In this limit, the defining equations (\ref{Y}) become:
\bea
x_1 x_2=0~~&,&~~x_3 x_4=0~~\nn\\
x_5 x_6=0~~&,&~~x_7 x_8=0~~,\nn
\eea\noindent so that $Y$ 
reduces to a union of $16$ copies of $\P^3$ intersecting with normal 
crossings\footnote{In this appendix, $Y_{\infty}$ and $Y_0$ mean 
$Y_{\psi=\infty}$ and $Y_{\psi=0}$, respectively.}:
\be
Y_\infty=\bigcup_{u_1,u_2,u_3,u_4\in \Z_2}{Z_{u_1,u_2,u_3,u_4}}~~,
\ee\noindent where $Z_{u_1,u_2,u_3,u_4}=
\{x=[x_1~...~x_8]\in \P^7~|~x_{1+u_1}=x_{3+u_2}=x_{5+u_3}=x_{7+u_4}=0\}\approx 
\P^3$. Following the procedure of \cite{Zharkov, Ruan1, Gross3, LV}, 
we consider the map
~$\mu:\P^7\longrightarrow \R^7$ given by:
\bea
\mu(x)=\frac{\sum_{k=1}^8{|x_k|^2P_k}}{\sum_{k=1}^8{|x_k|^2}}~~,
\eea\noindent with $P_1 ... P_8$ some points in general position in $\R^7$. 
The convex hull of these points defines a 7-simplex denoted by $\Delta$,
which clearly coincides with the image of $\mu$. According to the discussion of 
\cite{Zharkov, Ruan1, Gross3, LV}, a candidate for the desired  
$T^3$ fibration 
of $Y$ in the large radius limit is given by the restriction of 
$\mu$ to $Y_\infty$:
\be
\mu_0:=\mu|_{Y_\infty}:Y_\infty\rightarrow {\rm im}(\mu_0)\subset \Delta~~.
\ee\noindent Indeed, it is easy to see that the generic fiber of this map is 
a 3-torus. In the hypersurface case considered in 
\cite{Zharkov, Ruan1, Gross3, LV}, the image of $\mu_0$ coincides with the 
boundary of $\Delta$ (which is topologically a 3-sphere, since in the 
hypersurface case $\Delta$ has dimension 4), 
but for our complete intersection 
the situation is different. Indeed, it is easy to see that the image 
of each of the components $Z$ is a three-dimensional face of $\Delta$. For 
example, we have:
\be
\mu(Z_{1,1,1,1})=
\{\frac{|x_2|^2P_2+|x_4|^2P_4+|x_6|^2P_6+|x_8|^2P_8}{|x_2|^2+|x_4|^2+|x_6|^2+|x_8|^2}~|~(x_2,x_4,x_6,x_8)\in \P^3\}~~,
\ee\noindent which coincides with the three dimensional face 
$<P_2,P_4,P_6,P_8>$ spanned by the vertices 
$P_2, P_4, P_6$ and $P_8$. Hence the base of our fibration 
coincides with the union ${\rm im}(\mu_0)=\Delta_0$ of 16 
three-dimensional tetrahedra
\footnote{$\Delta_0$ is a subset of 
(but does not coincide with) the $3$-skeleton 
of $\Delta$ (i.e. the union of all of its three-dimensional faces).}. 
These tetrahedra intersect along common vertices, edges and facets, and the 
fibers of $\mu_0$ degenerate at the points of intersection.

What is the topology of the base $\Delta_0$ ? To answer this question, 
note that 
the 16 tetrahedra composing the base are spanned by the vertices:
\bea
\begin{array}{ccccccc}
<2,4,5,8>~&,&~ <1,4,5,8>~&,&~ <1,4,6,8>~&,&~ <2,4,6,8>\\
<2,4,5,7>~&,&~ <1,4,5,7>~&,&~ <1,4,6,7>~&,&~ <2,4,6,7>\\
<2,3,5,7>~&,&~ <1,3,5,7>~&,&~ <1,3,6,7>~&,&~ <2,3,6,7>\\
<2,3,5,8>~&,&~ <1,3,5,8>~&,&~ <1,3,6,8>~&,&~ <2,3,6,8>
\end{array}~~,\nn
\eea\noindent and $\Delta_0$ is obtained by gluing these along their common 
faces. Then a moment's thought shows that the resulting body is a 3-sphere
(see Figures 11 and 12). 
Thus, just as in the hypersurface case, 
$Y_\infty$ is a $T^3$ fibration over $S^3$.  

\vskip 0.5in 
\hskip 1.7in\scalebox{0.25}{\input{sphere1.pstex_t}}
\vskip 0.3in
\hskip 1in \begin{center}Figure 11. {\footnotesize  
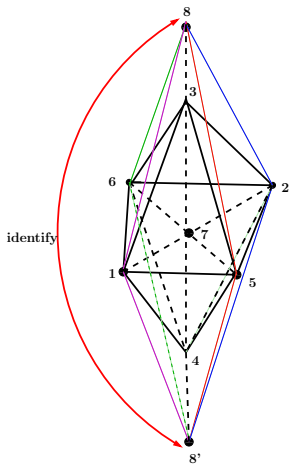
Arrangement of the 16 tetrahedra which form the base $\Delta_0$. The points 
$8$ and $8'$ are identified, together will all identifications of edges and 
facets implied by this.} 
\end{center}

\vskip 1.2in 
\hskip 1.0in\scalebox{0.2}{\input{sphere2.pstex_t}}
\vskip 0.3in
\hskip 1in \begin{center}Figure 12. {\footnotesize The 
identifications in Figure 11 can be performed in two steps. First, 
identify the edges starting from the point $P_8\equiv P_8'$; we represent 
this by introducing $6$ copies of that point. This shows that the topology of 
the base with the point $P_8$ removed is that of $\R^3$. 
Identifying the $6$ copies of $P_8$ amounts to adding a point to $\R^3$, 
which can be thought of as ``the point at infinity''. 
This produces a 3-sphere. }
\end{center}

\

Let us now consider the limit $z\rightarrow \infty\iff\psi\rightarrow 0$. 
In this limit, the defining equations reduce to:
\bea
x_1^2 +x_2^2=0~~&,&~~x_3^2+ x_4^2=0~~\nn\\
x_5^2+ x_6^2=0~~&,&~~x_7^2+ x_8^2=0~~,\nn
\eea\noindent which, via the transformation (\ref{change_coords}) are equivalent 
with:
\bea
\label{reduced_y}
y_1 y_2=0~~&,&~~y_3 y_4=0~~\nn\\
y_5 y_6=0~~&,&~~y_7 y_8=0~~.\nn
\eea\noindent Hence $Y_0$ reduces once again to 16 copies of $\P^3$ intersecting 
transversely, as should be expected from the fact that $Y_\psi$ 
and $Y_{1/\psi}$ are isomorphic as complex manifolds. Since the form of 
(\ref{reduced_y}) is the same as above, we can once again use the map:
\bea
{\tilde \mu}(y)=\frac{\sum_{k=1}^8{|y_k|^2 Q_k}}{\sum_{k=1}^8{|y_k|^2}}~~
\eea\noindent (with $Q_k$ some points in general position in $\R^7$) in order
to produce a $T^3$-fibration ${\tilde \mu}_0$ of $Y_0$ whose basis is a 
3-sphere. 

The fibrations $\mu_0$ and ${\tilde \mu}_0$ are related through
the biholomorphic map $\phi:Y_\infty\rightarrow Y_0$ 
which identifies the complex structures $J_\infty$ and $J_0$ 
of $Y_\infty$ and $Y_0$: 
\be
J_0=d\phi\circ J_\infty\circ (d\phi)^{-1}~~.
\ee\noindent We may hope that some appropriate deformations of 
the fibrations $\mu_0,~{\tilde \mu}_0$ are  
special Lagrangian with respect to $J_\infty,~J_0$ and the associated metrics.

\end{document}